\documentclass[usenatbib, useAMS]{mn2e}
\usepackage{psfig}
\usepackage{epsfig}
\usepackage{graphicx}
\usepackage{mathabx}
\usepackage{xspace}
\usepackage{color}
\usepackage [autostyle, english = american]{csquotes}
\MakeOuterQuote{"}

\def\avg#1{\left< #1 \right>}

\newcommand{\Euclid}{\textit{Euclid}\xspace}
\newcommand{\be}{\begin{equation}}
\newcommand{\ee}{\end{equation}}
\newcommand{\ba}{\begin{eqnarray}}
\newcommand{\ea}{\end{eqnarray}}

\newcommand{\mnras}{MNRAS}

\newcommand{\apjl}{Astro. Phys. Journal Letters}
\newcommand{\apj}{Astro. Phys. Journal}
\newcommand{\aap}{AAP}
\newcommand{\aj}{Astronomical Journal}
\newcommand{\apjs}{ApJS}             
\newcommand{\physrep}{Phys.~Rep.}   

\def\gs{\mathrel{\raise1.16pt\hbox{$>$}\kern-7.0pt %
\lower3.06pt\hbox{{$\scriptstyle \sim$}}}}         %
\def\ls{\mathrel{\raise1.16pt\hbox{$<$}\kern-7.0pt %
\lower3.06pt\hbox{{$\scriptstyle \sim$}}}}         %

\title[On Weak Lensing Shape Noise]{On Weak Lensing Shape Noise}

\author[]
       {Sami-Matias Niemi$^{1}$\thanks{email: s.niemi@icloud.com}, Thomas D. Kitching$^{1}$, Mark Cropper$^{1}$\\
       $^1$Mullard Space Science Laboratory, University College London, Holmbury St Mary, Dorking, Surrey RH5 6NT, UK}

\date{}

\pagerange{\pageref{firstpage}--\pageref{lastpage}}

\pubyear{2015}

\begin{document}

\maketitle

\label{firstpage}

\begin{abstract}
One of the most powerful techniques to study the dark sector of the Universe is weak gravitational lensing. In practice, to
infer the reduced shear, weak lensing measures galaxy shapes, which are the consequence of both the intrinsic ellipticity of the
sources and of the integrated  gravitational lensing effect along the line of sight. Hence, a very large number
of galaxies is required in order to average over their individual properties and to isolate the weak lensing
cosmic shear signal. If this `shape noise' can be reduced, significant advances in the power of a weak lensing
surveys can be expected. This paper describes a general method for extracting the probability distributions of parameters from
catalogues of data using Voronoi cells, which has several applications, and has synergies with Bayesian Hierarchical modelling approaches. This allows us to construct a probability distribution for the variance of the intrinsic ellipticity as a function of galaxy property using only photometric data, allowing a reduction of shape noise. As a proof of concept the method is applied to the CFHTLenS survey data. We use this approach to investigate trends of galaxy properties in the data and apply this to the case of weak lensing power spectra.
\end{abstract}


\section{Introduction}\label{Introduction}

Advances in observational cosmology in the last two decades have indicated that the expansion of the Universe is
accelerating \citep[e.g.][]{1998AJ....116.1009R, 1999ApJ...517..565P, 2010A&A...516A..63S, 2011ApJS..192...18K, 2013ApJS..208...20B, 2014A&A...571A..15P, 2014A&A...571A..16P, 2015arXiv150201589P}, rather than decelerating, as previously expected. Consequently, new models have been proposed as revisions to our current cosmological model (for a review see e.g. \cite{2013LRR....16....6A, 2012PhR...513....1C}), ranging from revisions of general relativity to the existence of a non-zero vacuum energy, generically termed ``dark energy'', to a revision in the fundamental large-scale assumptions for the Universe (such as isotropy and homogeneity, for papers discussing relaxing the assumption of isotropy, see e.g. \cite{2005ApJ...629L...1J, 2013MNRAS.436.3680M, 2014A&A...571A..26P, 2015arXiv150201593P}). Methodologies such as weak lensing and galaxy clustering are being developed to place observational constraints on these different approaches, and are becoming increasingly powerful, as they are explored and their characteristics and statistical behaviours understood.

One of the most powerful techniques currently available is weak gravitational lensing \citep[see e.g.][and references therein]{1992ApJ...388..272K, 2001PhR...340..291B, 2006astro.ph..9591A, 2006ewg3.rept.....P, 2008ARNPS..58...99H, 2008PhR...462...67M, 2010RPPh...73h6901M, 2013PhR...530...87W}. Through its gravitational effect, matter acts on light rays from a source to observer, and hence distorts the observed shape of the source; in analogy to an optical lens. Away from large concentrations of matter, such as the centre of galaxy clusters, this effect is very small: a change in the observed  third flattening, or third eccentricity (colloquially referred to as `ellipticity') of an image of a galaxy of only a few percent, an effect known as weak lensing. However if large enough samples of galaxies are observed, with sufficient attention to observational systematics, then a statistical analysis of the ensemble can be used to infer the matter distribution along the line of sight. Sources at higher redshift allow the mapping of the matter distribution at earlier times, and from the growth of structure and the relation of the amount of distortion to the expansion history of the Universe, inferences can be made on cosmological parameters \citep[e.g.][]{2006A&A...452...51S, 2007MNRAS.376..771K, 2007MNRAS.381..702B, 2008A&A...479....9F, 2010A&A...516A..63S, 2013MNRAS.430.2200K, 2014MNRAS.442.1326K}.

The change in the ellipticity of galaxies resulting from the gravitational lensing effect is known as `shear'.
Ellipticity is an unbiased shear estimator, however galaxies are also intrinsically (in the absence of lensing)
elliptical. However the orientation angle of the intrinsic ellipticity is expected to be randomly distributed on
large scales, which allows one to recover the shear by averaging over many galaxies.
In more detail, because galaxies to a first approximation have a random projected orientation on the sky the
intrinsic ellipticity acts as an additional noise term when attempting to measure the shear of an object: the
observed ellipticity is a sum of shear and intrinsic ellipticity $e=e^{\rm int}+g$, where $g$ is known as the
reduced shear. The variance on the observed ellipticity is $\sigma^2(e)=\sigma^2(e^{\rm int})+\sigma^2(g)$.
The variance of the intrinsic ellipticity is the dominant term, typically by a factor exceeding an order of magnitude.
Hence, a very large number of galaxies are required in order to reduce this by averaging over their individual properties and
to isolate the weak lensing signal. This term is commonly referred to as ``shape noise'',
and it is by far the largest source of uncertainty in weak lensing measurements.
Shape noise is a limiting fundamental noise source in weak lensing measurements that
determines the size of the observational surveys being designed to measure cosmological parameters using weak lensing:
future state-of-the-art surveys such as that to be carried out by the European Space Agency's {\it Euclid} mission \citep[for details, see]
[and \url{http://www.euclid-ec.org/}]{2011arXiv1110.3193L, 2012SPIE.8442E..0TL, 2013LRR....16....6A} encompass
most of the extragalactic sky. It follows, therefore, if this source of noise could be reduced, significant advances in the power of a weak lensing survey would follow, or smaller and less ambitious surveys could be contemplated while maintaining the previous survey sensitivity.

Recently \cite{2013arXiv1311.1489H} have noted that the Tully-Fisher relationship could in principle be used as a proxy for the intrinsic ellipticity of a galaxy. However Tully-Fisher analyses are not practicable in a weak
lensing context owing to the very large number of galaxies required, most of which are faint and hence have a low signal-to-noise ratio. Here we propose to generalise the concept to allow the incorporation of {\it any} other supplementary useful information that will provide prior knowledge on the intrinsic ellipticities, and which would be practicable. This could be any measurable property of a galaxy that does not have a correlation with shear\footnote{In this paper
we focus, as a proof of concept, on the ellipticity change in galaxies induced by weak lensing. The size and
magnification of galaxies is also affected by gravitational lensing, but for current surveys the impact of this
is expected to be small (see e.g. \cite{2014MNRAS.437.2471D}). However in extensions to this study the magnification effect should be taken into account.}.
In this paper we use as an example the optical magnitudes of the galaxies and their combination (pseudo-colours), in the expectation that certain galaxy types (characterised by certain colours) have a range of ellipticity which is different from other galaxy types (characterised by different colours) \citep[for a related study with COSMOS sample and the Millennium Simulation, in the context of intrinsic galaxy alignments, see][]{Joachimi01052013}.

What we propose is a general method to infer the probability distribution of the intrinsic galaxy ellipticity $p(|e|)$
from a weak lensing ellipticity catalogue, in particular how to characterise the variance of this distribution. We also show how the probability of this distribution can be accessed, as this is particularly useful in the context of Bayesian
hierarchical modelling \citep[e.g.][]{2014arXiv1411.2608S}. Currently a Gaussian probability distribution for $p(e_{i})$ is assumed, with a mean for the two ellipticity components $(i)$ of zero and a standard deviation $\sigma(e_{i})$. This standard deviation can be derived from an observed $p(|e|)$ from a calibration set (for example from higher spatial resolution subsets of the data), or it can be taken from the dispersion in the measurement of the ellipticities of the galaxies. If the ellipticity measurement is carried out using a model-fitting method, for example \emph{lens}fit \citep{Miller21112007}, the ellipticity distribution $p(|e|)$ also acts as prior, which again may be from a calibration set of higher resolution; as used in, for example the CFHTLenS analysis \citep{Miller11032013}.

The structure of this paper is as follows. In Section \ref{s:Method} we describe the data used and our methodology,
and then in Section \ref{s:Results} we show how the limitation of the resulting $p(|e|)$ ellipticity dispersion
width $\sigma(|e|)$ can be used to more optimally weight the dataset to achieve increased dark energy Figure of Merit (FoM)
\citep{2006astro.ph..9591A}.
Finally, we summarise and conclude in Section \ref{s:Conclusion}.

\section{Data and Methodology}\label{s:Method}

In this Section we present our general methodology. The central concept is that using clustering analysis (where the
proximity of data points as a function of input features is estimated) of catalogue data the functional dependency of the \emph{variance} of parameters of interest (in our case ellipticity of galaxies) on those parameters can be inferred. These
approaches in fact already have a far more general application than that which we propose here,
we are simply applying existing machine learning methodologies in a novel cosmological context. We first
describe the data that we will use to develop the example of this paper, before describing the general algorithm, and then the specifics
of the example implementation.

\subsection{Data}

The data we use are from the CFHTLenS \citep[see e.g.][]{2012MNRAS.427..146H} catalogue \citep{2013MNRAS.433.2545E}, which is a measurement of galaxy ellipticities (using the \emph{lens}fit method by \cite{Miller21112007}), galaxy photometric redshifts [\cite{2012MNRAS.421.2355H}, using BPZ \citep[e.g.][]{2000ApJ...536..571B, 2006AJ....132..926C}] and also includes in the catalogue derived meta-parameters such as galaxy stellar mass and bulge-to-disc ratio. The total data set corresponds to observations in five optical bands, \textit{ugriz}, over $154$ square degrees of sky. We use the catalogues downloaded from \url{http://www.cadc-ccda.hia-iha.nrc-cnrc.gc.ca/community/CFHTLens/query.html}, and make no cuts in the catalogue.

\subsection{Algorithm Overview}

The algorithm, expanded upon in detail in the following sections, is in general terms the following. We assume that there are some parameters of interest (in our case ellipticity), and some meta-parameters that may (or may not) be related to the parameters of interest. The objective is to determine how the probability distribution of the parameters of interest is a function of the meta-parameters. Using a typical catalogue the dependency of the mean of the parameters is trivial to determine, but the full probability distribution is not.

In summary, we adopt the following approach:

\begin{enumerate}
\item First we perform a dimensionality reduction on the data. This can be of any form, but is required to remove degeneracies and to enable computationally efficient subsequent steps. In our example we use principal component analysis \citep[PCA;][]{doi:10.1080/14786440109462720}, but this could be generalised for example to  manifold learning (e.g. locally-linear embedding, isomap, etc.; \cite{2012arXiv1206.5538B}) or auto-node mapping (where the input is linked to the output nodes) in neural net analysis. At this stage a removal of the covariance of the parameters is also beneficial.

\item Next we perform a clustering analysis on the resulting parameters. This clustering analysis should be designed to find points in the parameters that are `close' in some pre-defined distance measure (the most straightforward being the geometric distance between the points in the parameter space, but more general measures can be envisioned).

\item Finally the statistical properties in each cluster cell, labelled by a combination of meta-parameters, of the parameters of interest can now be computed. For example, this could be the mean and standard deviation, amongst other statistics. Furthermore, probability density functions can also be derived using, for example, kernel density estimation \citep[see e.g.][]{rosenblatt1956, parzen1962, Rudemo1982, Cao94}.
\end{enumerate}

We now specify the details of the implementation of this approach that we take as an example in this paper, and apply this methodology to the CFHTLenS catalogue. We then apply this methodology to the case of reducing the shape noise term in weak lensing power spectrum analysis in an effort to increase the accuracy on the measurement of cosmological parameters for weak lensing analysis.

\subsection{Dimensionality Reduction}\label{ss:dimensionalReduction}

Our chosen technique relies on identifying relations between photometric data and ellipticity measurements. The clarity of these relationships will be more evident if we project the five dimensional space of the CFHTLenS photometry (from the five bands) to a lower-dimensional surface within it which captures most of the information content. Our motivation for
dimensionality reduction is to find those directions in the data that contain the majority of the information on the correlation (of more general relationship) between variables. Hence we perform a dimensionality reduction using PCA \citep{doi:10.1080/14786440109462720}, which uses an orthogonal transformation to convert a set of observations of possibly correlated variables into a set of values of maximally linearly uncorrelated orthogonal variables called principal components. Before performing the dimensionality reduction with PCA we whiten\footnote{A whitening changes an input vector into a white noise vector by transforming a set of random variables with a known covanriance matrix into a set of new random variables which are uncorrelated and all have variance equal to unity. In practice, the vectors are divided by the singular values.} the data allowing to project the data onto the singular space while scaling each component to unit variance. This is useful because the clustering algorithm we adopt below makes assumptions on the isotropy of the signal. In the analyses we use the \textit{RandomizedPCA} \citep{2009arXiv0909.4061H} implementation of the Scikit Learn package \citep{scikit-learn}. In this process, the linear combinations of magnitudes effectively produces a colour-colour hyperspace.

\begin{figure}
\includegraphics[width=84mm]{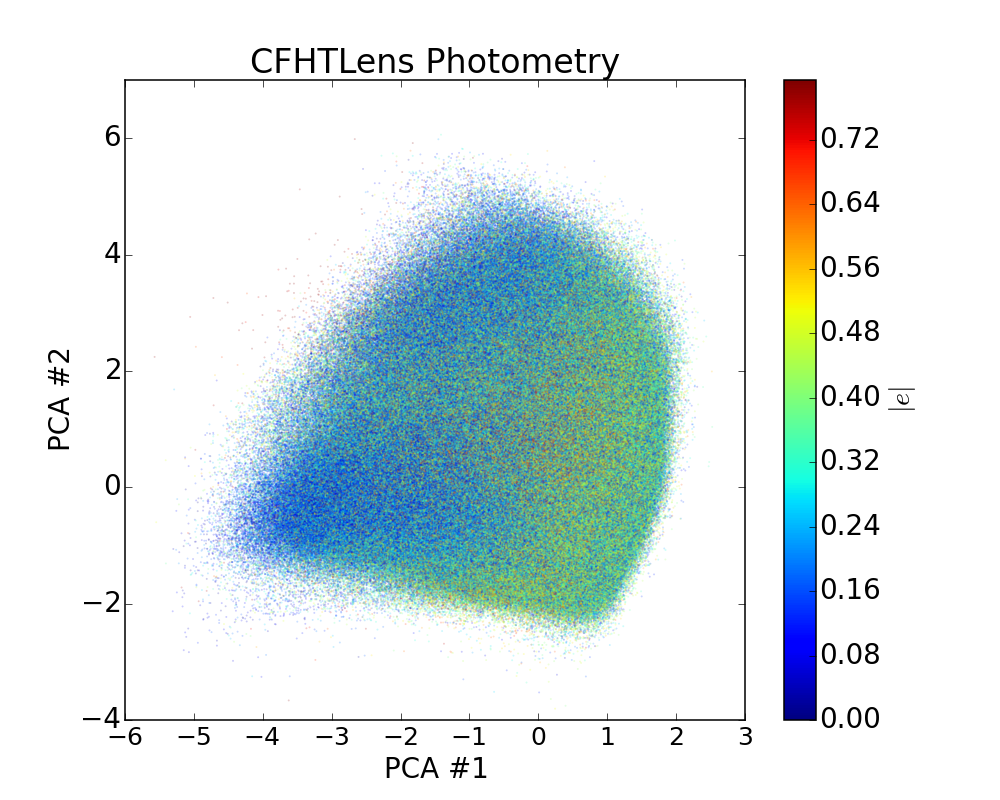}
\caption{CFHTLenS photometry and colours projected onto two dimensions using dimensionality reduction by the PCA method. Each galaxy has been colour coded by its measured ellipticity $|e|$. A low value of PCA\#1 corresponds to high apparent brightness.}
\label{fig:PCA}
\end{figure}

Figure \ref{fig:PCA} shows the CFHTLenS photometry projected onto the two dimensions with the highest eigenvalue
principal components. Jointly the two first principal components explain approximately $96$ per cent of the variance in the
original CFHTLenS photometry.  The PCA components with the two highest eigenvalues are the following:
\begin{equation}\label{eq:PCAvectors}
\left( \begin{array}{rrrrr}
0.20 &  0.22 &  0.25 &  0.25 &  0.25 \\
0.96 &  0.49 & -0.08 & -0.45 & -0.67 \\
\end{array} \right)
\left( \begin{array}{c}
u\\
g\\
r\\
i\\
z\\
\end{array} \right) \, .
\end{equation}
Thus, PCA\#1 can be interpreted to represent an averaged broad band (`white light') magnitude, while PCA\#2 amounts to a colour term `blue $-$ red', where the blue part is a combination of $u$ and $g$ and the red is comprised mostly of $i$ and $z$.

In Figure \ref{fig:PCA}, we colour-code each of the $5770490$ galaxies with the modulus of their ellipticity $|e|$,
calculated from the ellipticity components $e_{i}$ as $|e| = \sqrt{e_{1}^{2} + e_{2}^{2}}$, measured in the CFHTLenS survey. Regions with lower mean ellipticity, broadly, those with negative values in the first two principal components are evident. However, to quantify the mean ellipticities of galaxies in the different regions, we must create samples which have similar values in the PCA\#1 -- PCA\#2 plane.

\subsection{Clustering}\label{ss:clustering}

While the simplest grouping of objects can be achieved by simply binning in multiple dimensions (by grouping objects based
on their positions), this is not optimal when the object density varies in the PCA\#1 -- PCA\#2 plane, because some cells are more
densely populated while others are sparse. It would be better to divide the space into regions with similar photometric parameter
ranges. This can be done in a number of different ways, for example by a self-organising map \citep{KohonenMap}. Here we instead adopt a k-means \citep[e.g.][]{zbMATH03129892, zbMATH03340881} clustering method (abbreviated to `k-means'), because self-organising maps can be viewed as a constrained version of k-means clustering \citep{hastie01statisticallearning} and we aim to propose a general methodology.

It can be shown \citep[see e.g.][]{kmeans1, kmeans2} that PCA automatically projects to a subspace where the global solution of k-means clustering lies, and thus facilitates k-means clustering to find near-optimal solutions. It therefore is natural to adopt k-means, over self-organised maps, to identify regions of similar feature qualities in the projected space. k-means aims to partition
the galaxies into $k$ clusters in which each galaxy belongs to the cluster with the smallest sum of squares distance in the feature space. This results in a partitioning of the projected data space into Voronoi cells. In the analyses we use the \textit{MiniBatchKMeans} implementation of the Scikit Learn package \citep{scikit-learn}.

\begin{figure}
\includegraphics[width=84mm]{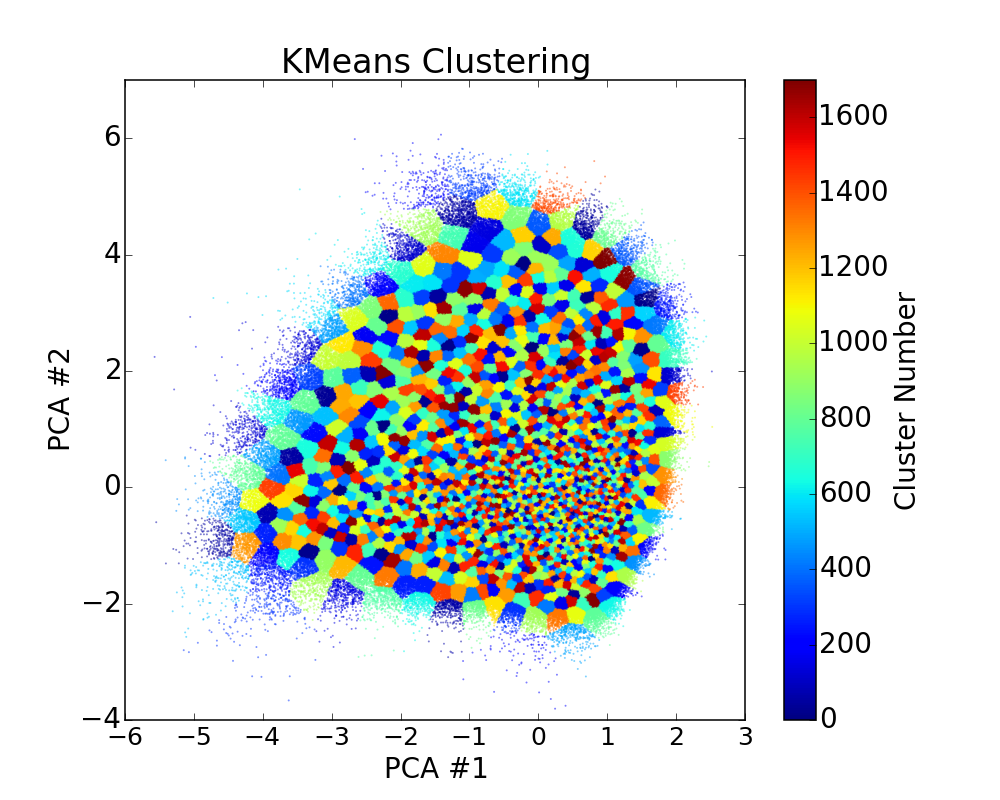}
\caption{Clusters identified by k-means algorithm from the PCA projected data. The projected PCA plane has been partitioned to 1700 Voronoi cells. A low value of PCA\#1 corresponds to high apparent brightness.}
\label{fig:kmeans}
\end{figure}

Figure \ref{fig:kmeans} shows the results of k-means clustering when the galaxies are grouped into $(\sqrt{5770490/2}) \sim 1700$ clusters. Because there are more galaxies towards the centre of the PCA plane, the Voronoi cells there are smaller compared to the
cells around the edges. This achieves the objective of a more even number of galaxies in each cell and hence smaller
mean-level differences between the cells. 

\subsection{Statistics of Parameters of Interest: Ellipticities}\label{ss:ellipticities}

Now that we have implemented the algorithm described above we can begin to investigate the behaviour of the parameters of interest, in our case the ellipticities of the galaxies used in the weak lensing analysis. In the following the mean and the standard deviation of the ellipticities refer to the weighted versions of these quantities:
\begin{eqnarray}
\avg{e_{i}}  & = & \frac{\sum w_{i} e_{i}}{\sum w_{i}}\label{eq:weighted_e} \\
\sigma(e_{i}) & = & \frac{\sum w_{i}(e_{i} - \avg{e_{i}} )^{2}}{\sum w_{i}} \quad .
\end{eqnarray}
The weight $w_{i}$ for a given galaxy $i$ is taken directly from the CFHTLenS catalogue and refers to the inverse variance as calculated by the \textit{lens}fit algorithm.

\subsubsection{Mean}

\begin{figure*}
\includegraphics[width=\columnwidth]{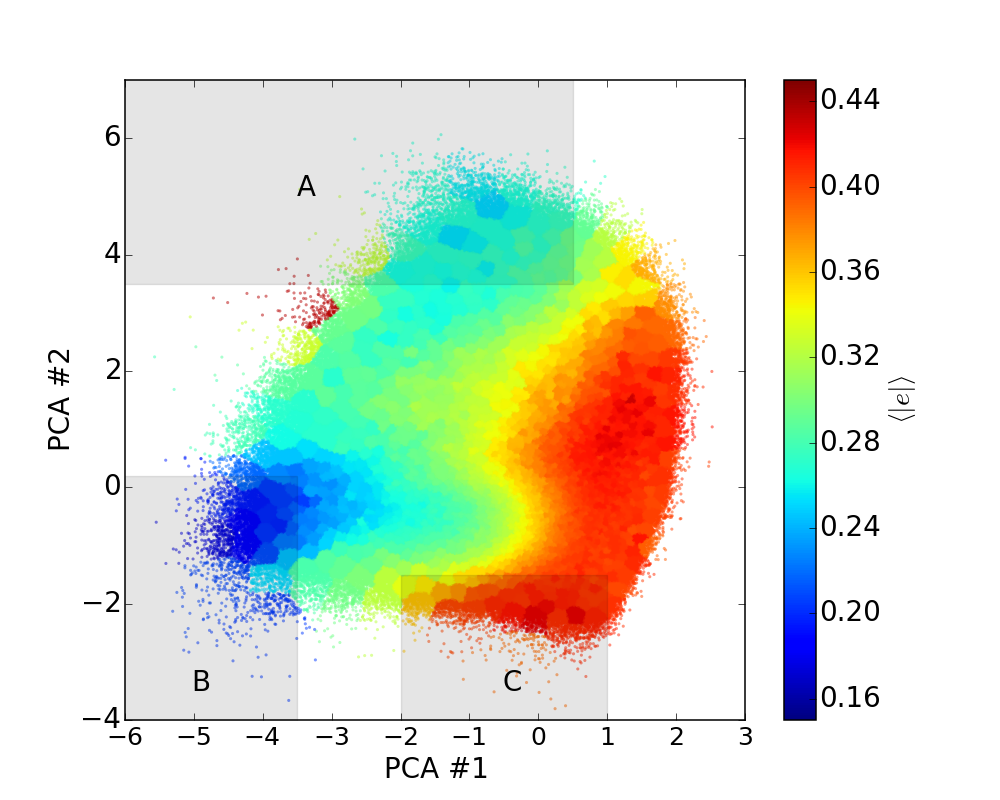}
\includegraphics[width=\columnwidth]{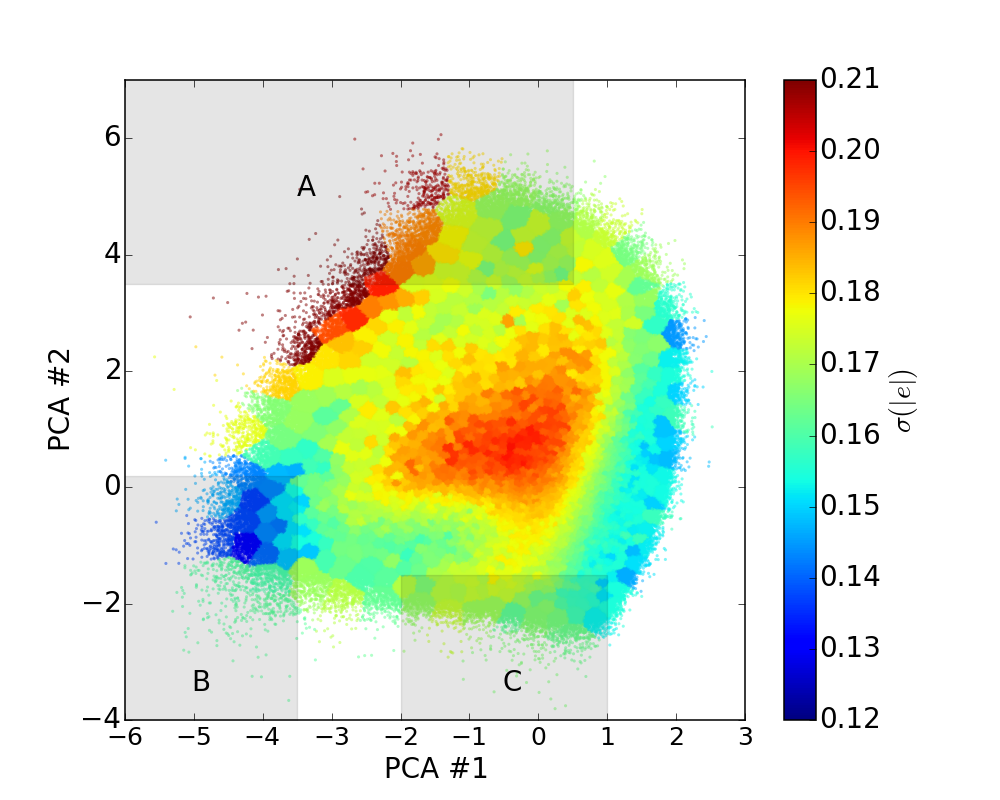}
\caption{Left: the mean ellipticity $\avg{|e|}$ in each of the 1700 Voronoi cells identified by k-means algorithm from the PCA projected data. Right: the standard deviation of modulus of ellipticity $\sigma(|e|)$ of the galaxies in each of the 1700 cells identified by k-means algorithm from the PCA projected data. Three regions of interest, named "A", "B", and "C", have been highlighted. A low value of PCA\#1 corresponds to high apparent brightness.}
\label{fig:kmeansEllipticity}
\end{figure*}

The left panel of Figure \ref{fig:kmeansEllipticity} shows the mean of the modulus of ellipticity in each of the $1700$ Voronoi cells in Figure \ref{fig:kmeans} identified by the k-means algorithm from the PCA projected data. The colour coding shows low and high mean ellipticity regions with blue and red, respectively. As a sanity check of the data we also calculated the mean of each of the ellipticity component $\avg{e_{1}}$ and $\avg{e_{2}}$, which should be consistent with zero in the absence of systematic effects. These were found to be consistent with zero (all Voronoi cells have $-0.015 \leq \, \avg{e_{i}} \, \leq 0.015$).

It is remarkable how strongly the average of the modulus of the ellipticity $\avg{|e|}$ is correlated with the PCA\#1 component of the photometry, with a lesser dependence on PCA\#2. Given that PCA\#1 relates to the visual brightness of a galaxy, this can be explained by three observations. Firstly, at lower redshifts $(z \ls 1)$ intrinsically brighter objects are likelier to be more circular from galaxy evolution arguments \citep[e.g.][]{1996AJ....112..359V, 2000ApJ...529..886C}: such objects are generally elliptical galaxies instead of spirals, and ellipticals on average have a lower measured $|e|$. Secondly, it is also possible that the galaxies in the low mean ellipticity Voronoi cells are mostly bright face on spiral galaxies, which tend to be less elliptical and visually brighter than more elliptical edge on spirals, which may suffer from substantial dust attenuation in optical wavelengths. In addition, a third effect will arise from what has been termed as `noise bias' \citep[][and references therein]{2014MNRAS.439.1909V} in the weak lensing community. That is, the measured ellipticity of any object is more uncertain (and in general biased towards zero) as the noise level in an image increases. In practice, it is likely that all three competing effects contribute.

\subsubsection{Standard Deviation}

The right panel of Figure \ref{fig:kmeansEllipticity} shows the standard deviation of the ellipticities of the galaxies in each of the $1700$ Voronoi cells. The figure shows that the region with the highest $\sigma(|e|)$ is at the centre of the projected parameter space and also at the top left corner. Interestingly, the highest mean ellipticity does not correspond to the highest standard deviation regions in the colour-colour principal components plane. The highest ellipticity variance corresponding to PCA\#1 values of $\sim 0$ is a significant feature, rather than simply a noise artefact, as we explain below.

\begin{figure*}
\includegraphics[width=52mm]{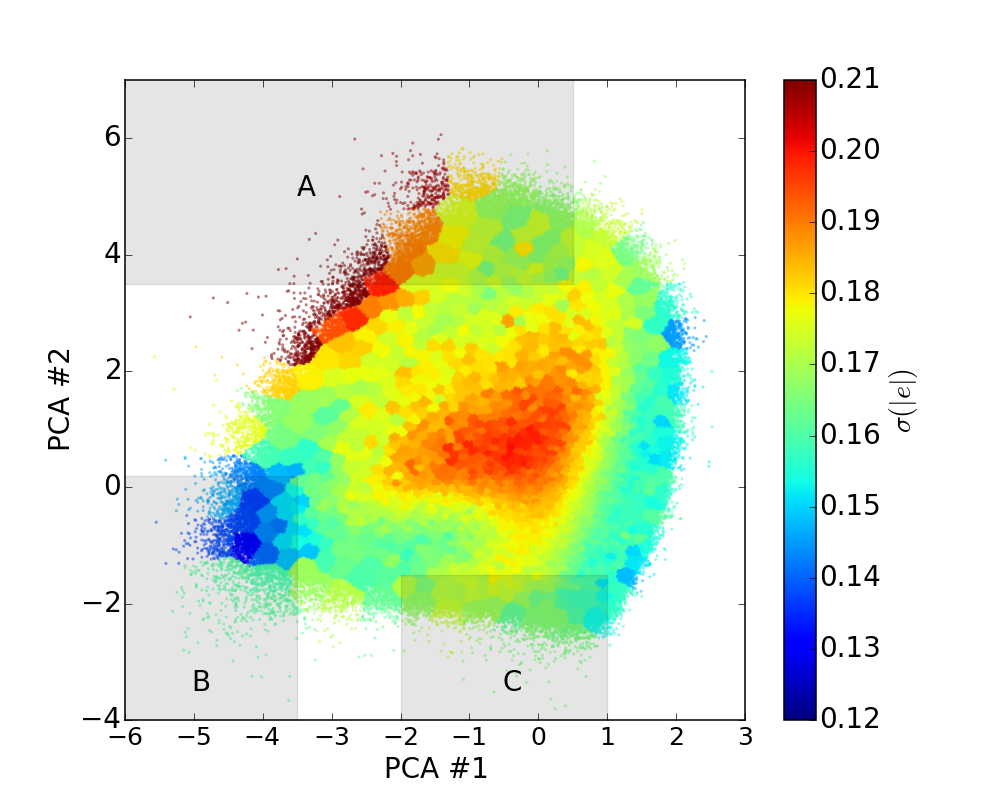}
\includegraphics[width=52mm]{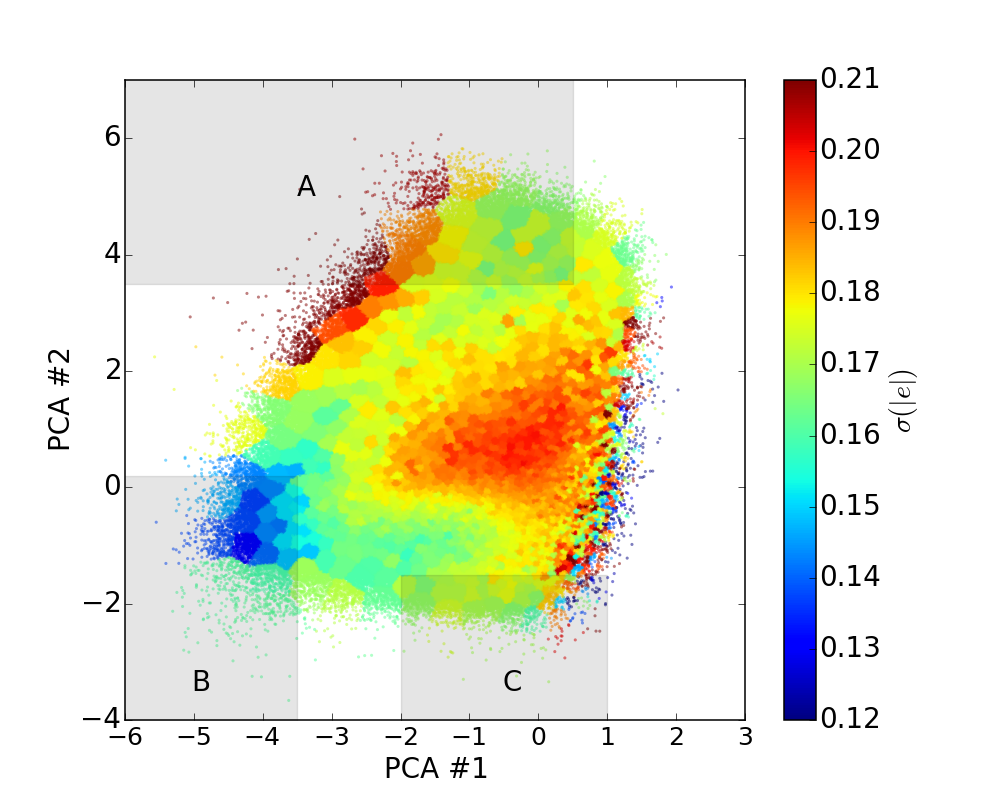}
\includegraphics[width=52mm]{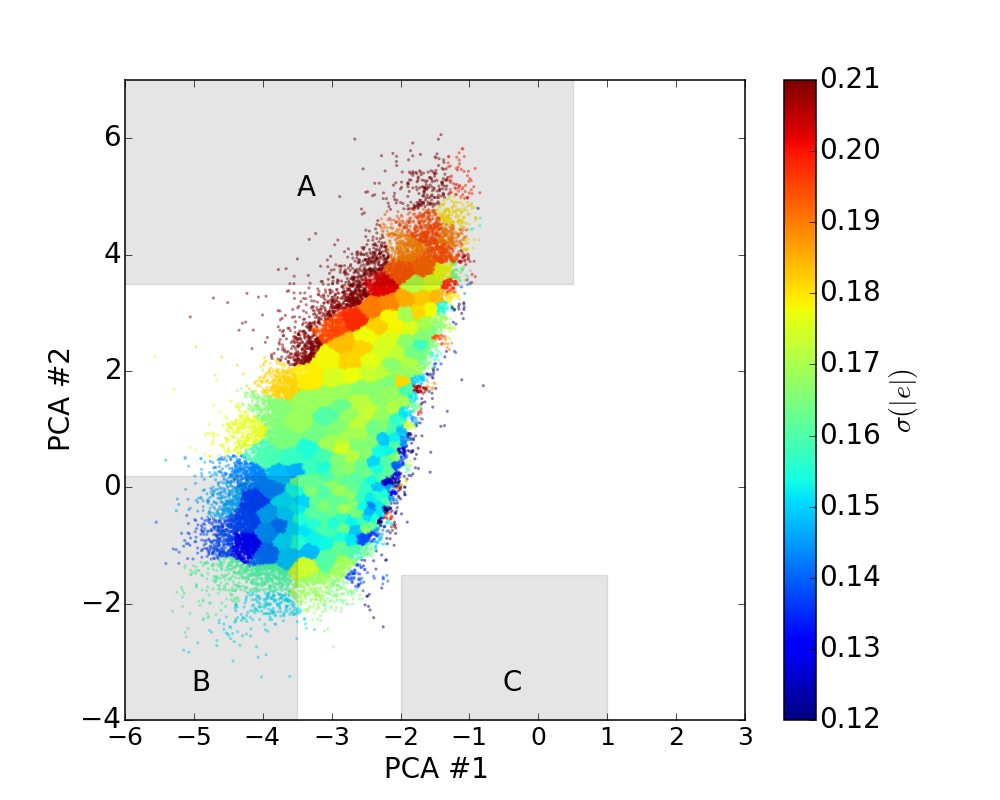}
\caption{The standard deviation of modulus of ellipticity $\sigma(|e|)$ of the galaxies in each of the 1700 cells identified by k-means algorithm from the PCA projected data with different signal-to-noise ratio cuts Left: All galaxies with SNR $> 10$. Middle: All galaxies with SNR $> 20$. Right: All galaxies with SNR $> 100$. A low value of PCA\#1 corresponds to high apparent brightness and a low value of PCA\#2 to a bluer galaxy colour.}
\label{fig:SNRexample}
\end{figure*}

In some cases the high standard deviations could result from larger measurement uncertainties, rather than the intrinsic dispersion. However, in the following we show that the standard deviations over a large part of the PCA plane, especially at high PCA\#1 values with the faintest galaxies, are not dominated by measurement errors, but rather the variations in the measured galaxy ellipticities. While the ellipticity measurement for an individual galaxy becomes more uncertain as we move along the PCA\#1, simply because the signal-to-noise ratio decreases as the PCA\#1 increases, we do not observe the highest dispersions with the highest PCA\#1 values. While the noise contribution is a monotonic function of PCA\#1, its contribution should not \emph{exceed} the variance of the cells at a given PCA\#1. For all Voronoi cells with the centre of their PCA\#1 greater than $1.3$, we find the standard deviation to be $\sim 0.004$ in the ensemble of cells for $\sigma(|e|)$ compared to the typical values of $\sigma(|e|)$ of $\sim 0.14$. Hence, even at the faintest PCA\#1 the broadening by measurement noise is not dominant, and it is even less so in all other cells, by thus setting an upper limit for the broadening by measurement noise. To demonstrate this further we have also repeated the analysis at different signal-to-noise ratio slices. The main findings of the paper are largely independent from the selected SNR. Most importantly, the cell statistics remain unchanged when adopting a different SNR cut, only the number of cells populated by galaxies change, as is evident from Figure \ref{fig:SNRexample}. In the three panels, the SNR selections range from $>10$, $>20$, and $>100$. For the cells that are well populated, the mean ellipticity and the standard deviation remain unchanged.

\begin{figure*}
\includegraphics[width=\columnwidth]{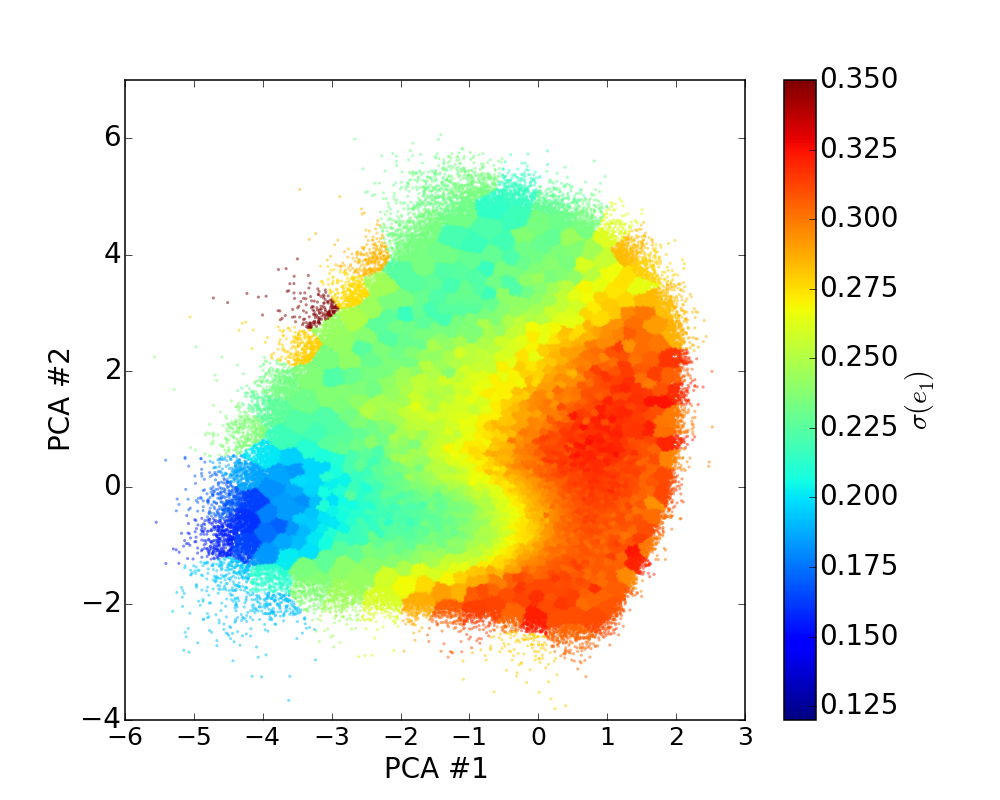}
\includegraphics[width=\columnwidth]{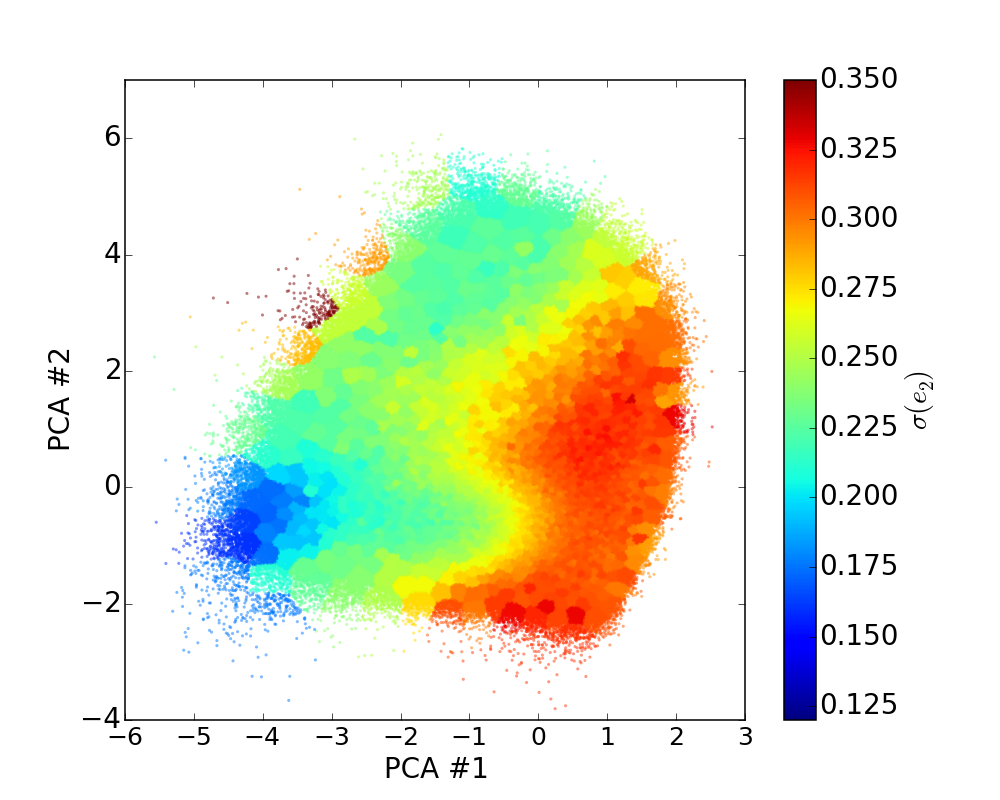}
\caption{Left: the standard deviation of the first ellipticity component $\sigma(e_{1})$ of the galaxies in each of the 1700 cells identified by k-means algorithm from the PCA projected data. Right: the standard deviation of the second ellipticity component $\sigma(e_{2})$ of the galaxies in each of the 1700 cells identified by k-means algorithm from the PCA projected data. A low value of PCA\#1 corresponds to high apparent brightness and a low value of PCA\#2 to a bluer galaxy colour.}
\label{fig:sigmaEs}
\end{figure*}

While the right panel of Figure \ref{fig:kmeansEllipticity} shows the standard deviation of the modulus of the ellipticity, however, for weak lensing analysis the standard deviations of each of the ellipticity components is also of interest. We therefore show the standard deviations for both components separately in Figure \ref{fig:sigmaEs}. It is evident from the Figure that these agree well with each other and also, to some extent, with the mean of the modulus of the ellipticity (the right panel of Figure \ref{fig:kmeansEllipticity}). At high signal to noise we find a similar behaviour, whereas at low signal to noise we find some deviation, where the individual-component variance rises, where the modulus' variance decreases. This results from two effects that are more pronounced at low signal to noise.  Firstly, the standard deviation of the modulus (which is always positive semi-definite), which is approximately a Rayleigh distribution over $0 < |\,e\,| < 1$, is smaller than the standard deviation that one would compute from a Gaussian over the interval $-1 < e_{i} < 1$, the relation being a factor of $(4-\pi)/2$. Secondly $e_{1}$ and $e_{2}$ are correlated, thus the standard deviation of the individual components is not expected to fully trace the variance of the modulus. In the remainder of this paper we will investigate the modulus although these considerations should be kept in mind in the interpretation of the results.

We can also identify broad regions of interest in the PCA plane, these are named "A", "B", and "C", in Figure \ref{fig:kmeansEllipticity}. The "A" region, located at high $(\gs 3.5)$ PCA\#2  and intermediate $(\ls 0.5)$ PCA\#1 values, contains Voronoi cells with low or average mean ellipticity, but medium to high standard deviation. Instead, the region "B", found with the smallest PCA values (PCA\#1 $\ls -3.5$ and PCA\#2 $\ls 0.2$), contains Voronoi cells with the lowest mean ellipticity and standard deviation. This is especially interesting for weak lensing. The third region, named "C", can be found with the smallest $(\ls -1.5)$ PCA\#2 and intermediate $(-2 \ls \textrm{PCA\#1} \ls 1)$ PCA\#1 values, and contains Voronoi cells with the largest mean ellipticity and intermediate standard deviation. It will be interesting to see what types of galaxies reside in these cells. Finally, the rest of the area is covered mostly by cells with either large mean or large standard deviation. We will return to these regions in Section \ref{ss:regionsInterest}.

\begin{figure*}
\includegraphics[width=52mm]{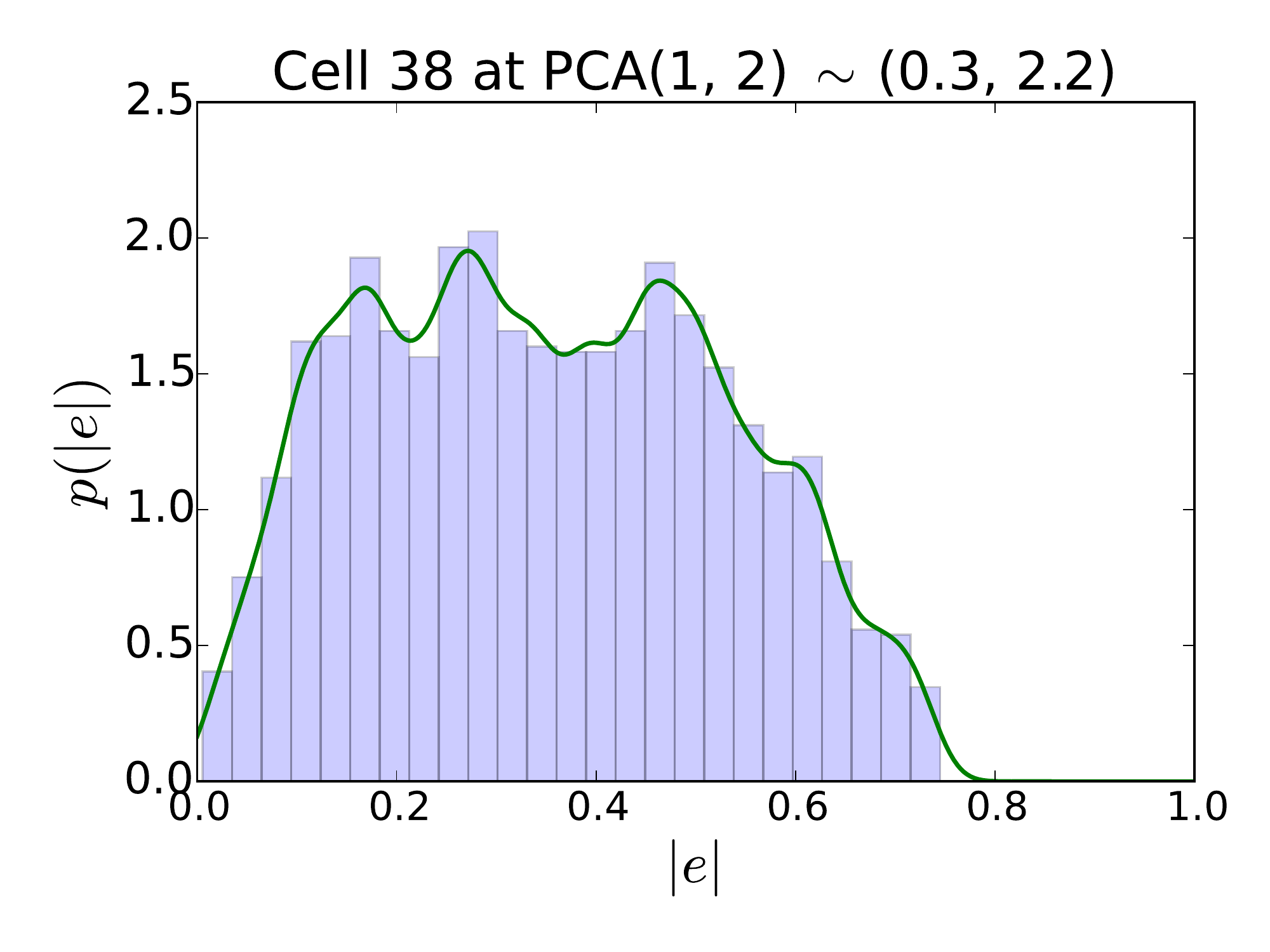}
\includegraphics[width=52mm]{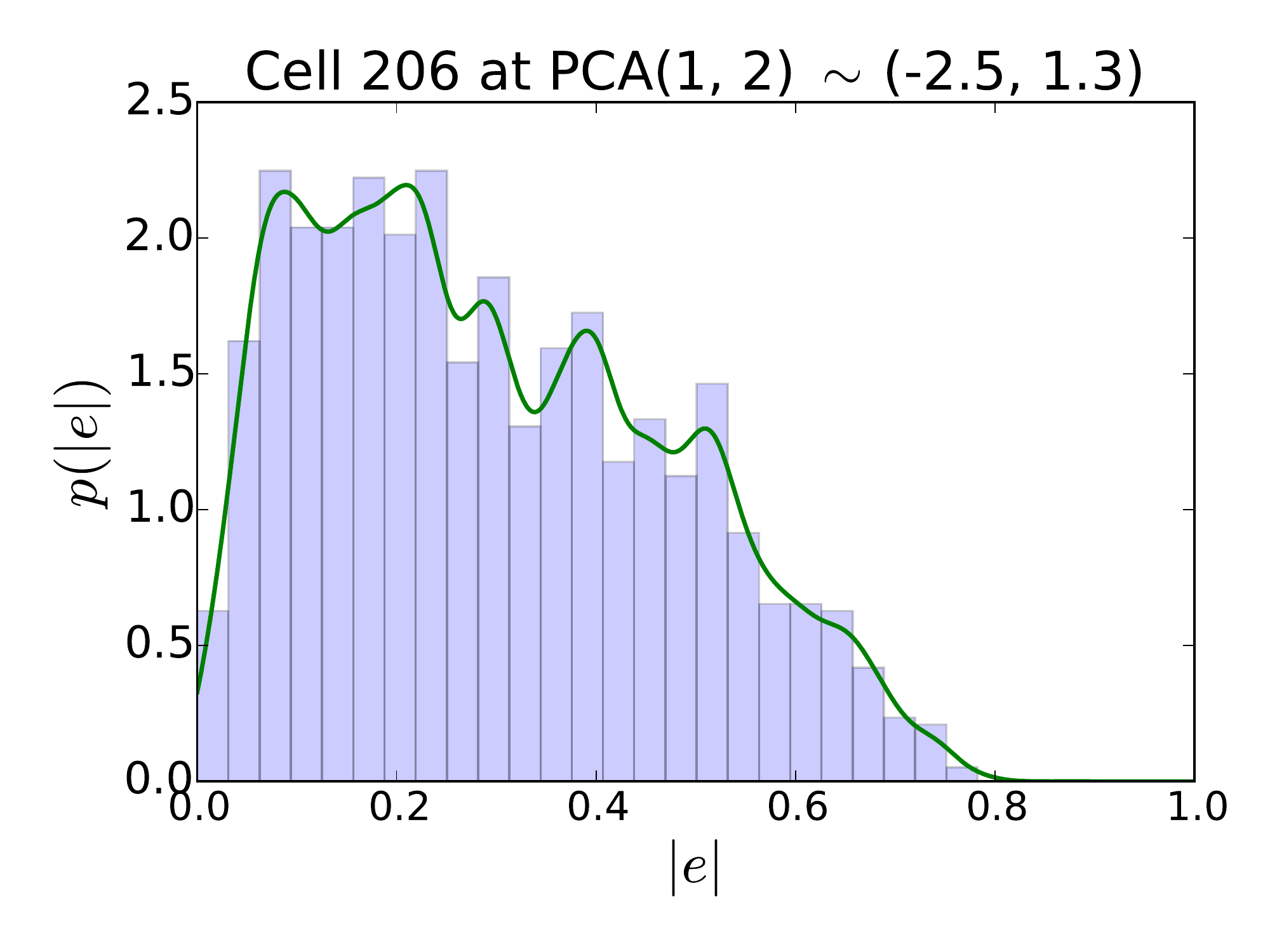}
\includegraphics[width=52mm]{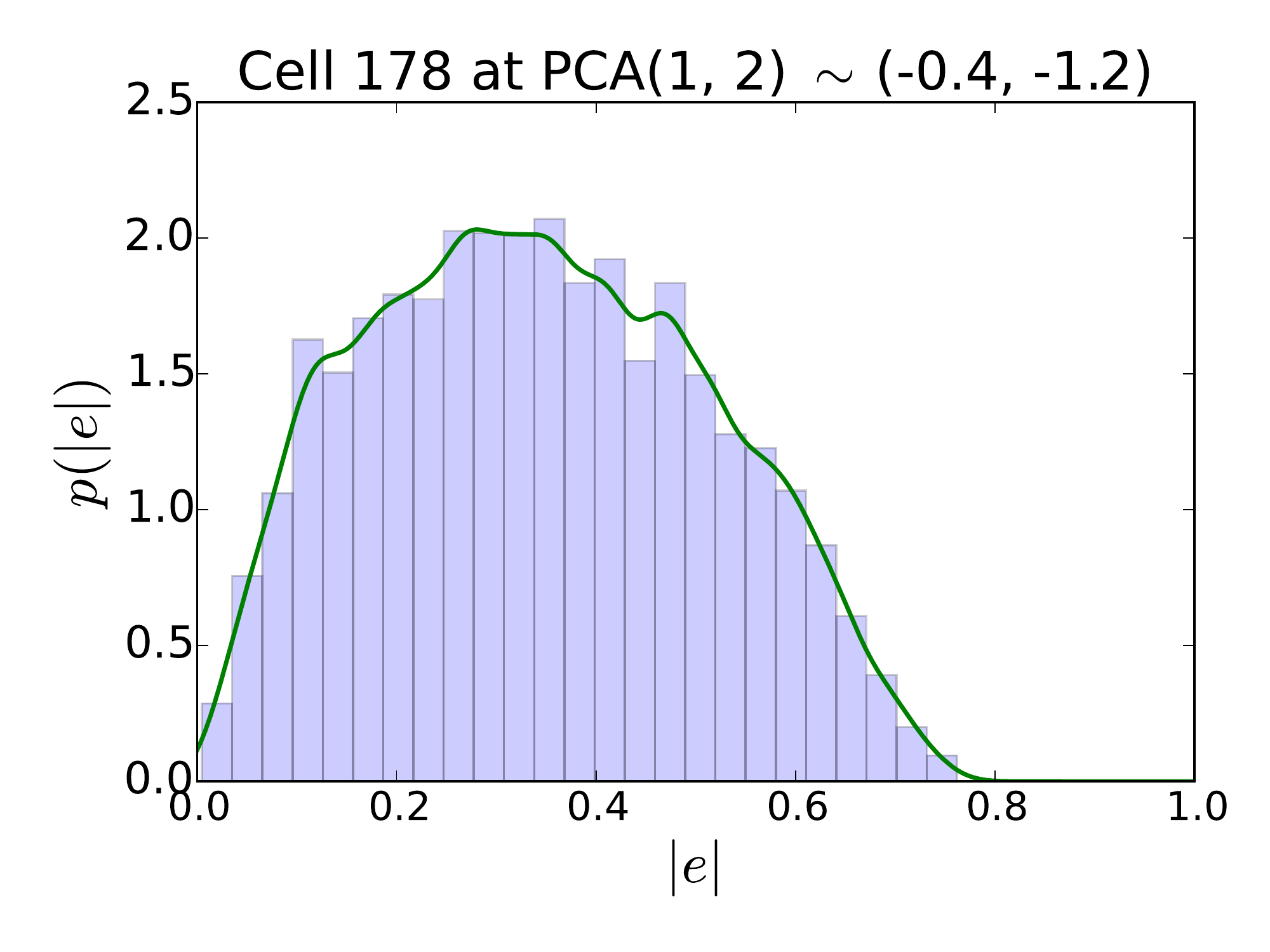}
\caption{Left: Probability density function for the weighted  ellipticity from the galaxies of cell 38 containing mostly very massive elliptical galaxies at intermediate redshifts (region "A"). Middle: Probability density function for the weighted ellipticity from the galaxies of cell 206 containing mostly face-on disk or $S0$-galaxies at relatively low redshift (region "B"). Right: Probability density function for the weighted ellipticity from the galaxies of cell 178 containing mostly inclined disk galaxies at high redshift (region "C").}
\label{fig:cellPDFs}
\end{figure*}

\subsubsection{Probability Distributions of the Cells}

Figure \ref{fig:cellPDFs} shows the probability density functions (PDFs) for ellipticity derived from all galaxies in a
given cell for three different Voronoi cells, corresponding to the three regions described above.  These are representative examples. 
The left panel of Fig. \ref{fig:cellPDFs} shows the ellipticity PDF of cell number $38$ in Region A, which contains mostly very massive elliptical galaxies at intermediate redshifts. It is clear that the Voronoi cell must also contain some disk or $S0$-galaxies that are inclined because the tail of the PDF extends to higher ellipticities. The middle panel shows the ellipticity PDF of cell number $206$ in region B, which contains mostly face-on disk or $S0$-galaxies at relatively low redshift, peaks at low ellipticity values with a skewness to higher ellipticity values. This will lead to a low mean cluster ellipticity and a relatively small variance. The right panel of Fig. \ref{fig:cellPDFs} shows the ellipticity PDF of cell number $178$ in Region C, containing mostly inclined disk galaxies at high redshift. The peak of the PDF has now shifted to intermediate ellipticities and the distribution is rather broad leading to a relatively high variance. These panels can be compared to Figure 7 of \cite{Joachimi01052013}. In general, we find a good agreement. For example, the top panel of Figure 7 of  \cite{Joachimi01052013} showing the PDF for late-type galaxies is in good qualitative agreement with the right panel of Fig. \ref{fig:cellPDFs}.

\subsubsection{Probability Distributions}\label{ss:ePDF}

\begin{figure*}
\includegraphics[width=79mm]{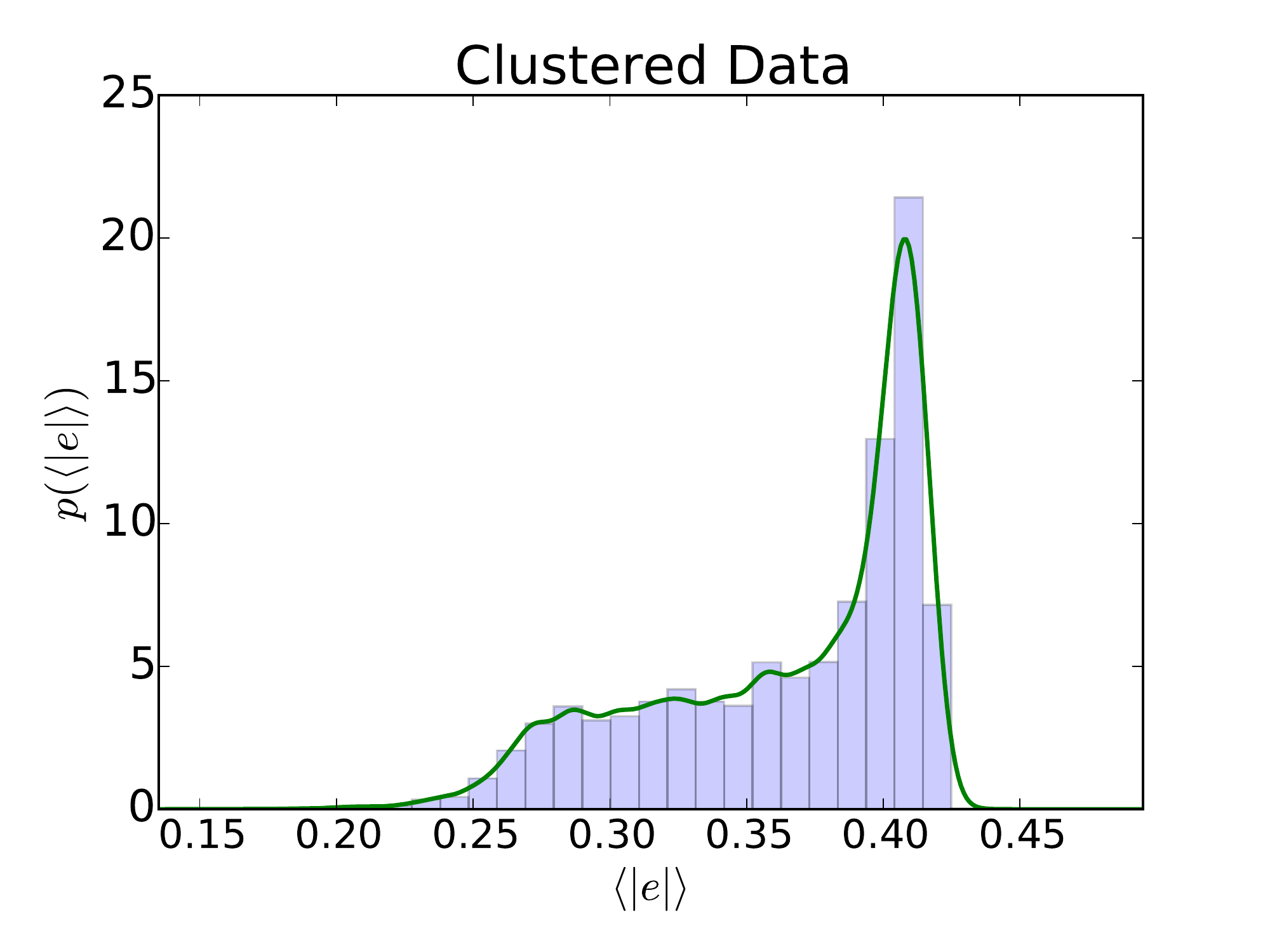}
\includegraphics[width=79mm]{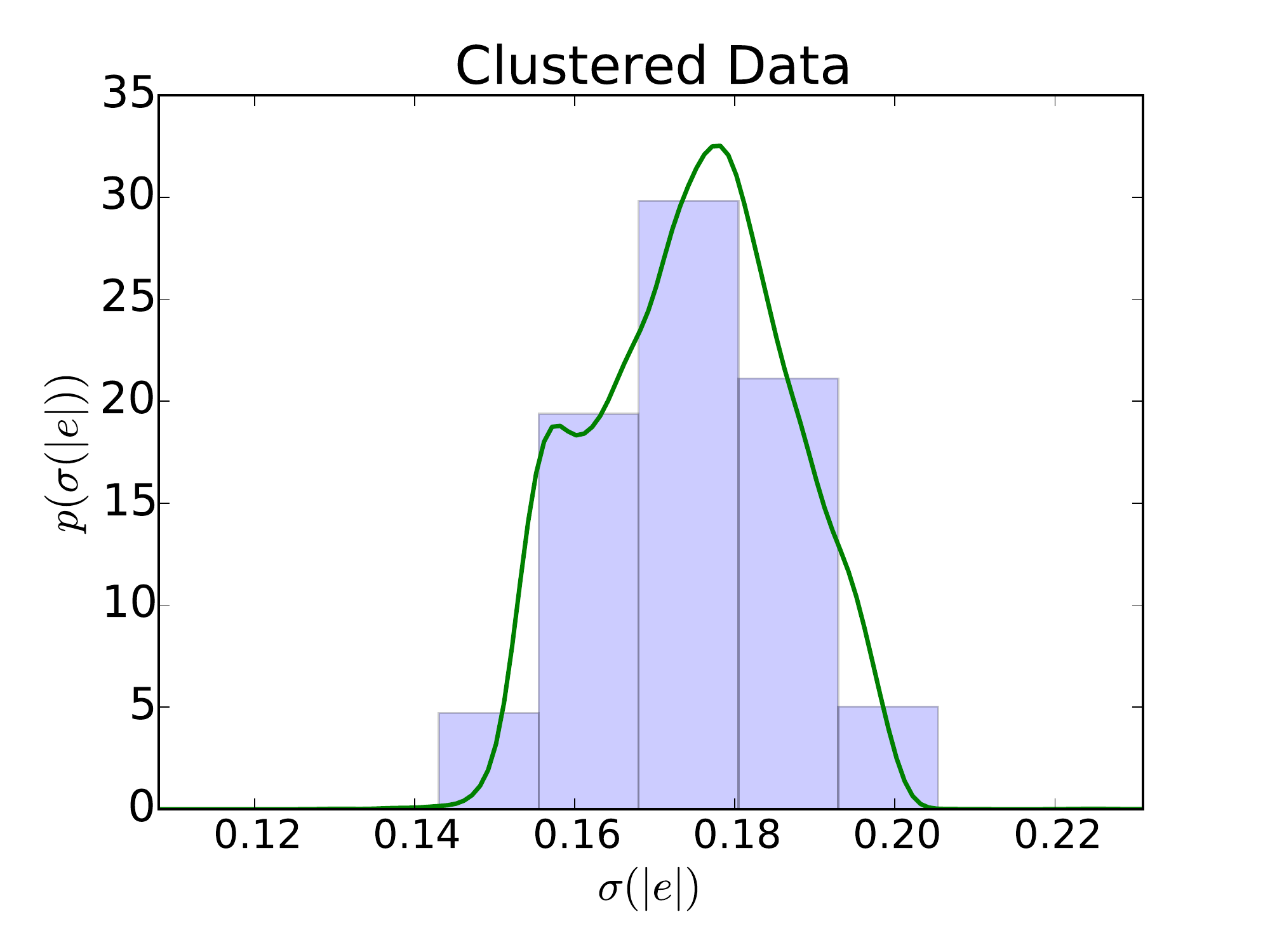}
\caption{Left: probability density function for the weighted mean ellipticity $|e|$ from the clustered data with 1700 individual Voronoi cells. Right: probability density function for the weighted standard deviation of the ellipticity $\sigma(|e|)$ from the clustered data with 1700 individual cells.}
\label{fig:PDFs}
\end{figure*}

The left panel of Figure \ref{fig:PDFs} shows the ellipticity probability density function derived from all $1700$ Voronoi cells in
Figure \ref{fig:kmeans}, while the right panel of Figure \ref{fig:PDFs} shows the probability density function of the standard
deviation of the ellipticities $p(\sigma(|e|))$. Note that these PDFs (from all cells) are different from those in Fig. \ref{fig:cellPDFs} (constructed from galaxies within a cell). By selecting galaxies with particular projected PCA values, we can select the appropriate $p(\sigma(|e|))$. In the following Sections we apply this to the case of a weak lensing power spectrum analysis. For example, the $\sigma_{\alpha}(|e|)$ can be tailored to select those particular galaxies (denoted here with $\alpha$) that may be  narrower than the all-encompassing $\sigma(|e|)$ for all galaxies. Therefore the shape noise in the power spectrum calculations can be reduced. This will be pursued in Section \ref{s:Results}.

\subsection{Redshift}\label{ss:redshifts}

The weak lensing power spectrum must be generated at different redshifts to enable weak lensing tomography,
as this requires several redshifts to measure the dark energy equation of state (dark energy being an
effect that changes the expansion history of the Universe as a function of redshift). It is therefore interesting to look at the redshift distribution of galaxies measured in the CFHTLenS survey in our k-means Voronoi cells.  We use the probability density function estimates (full posterior redshift information) from CFHTLenS catalogues, but visualise the point estimate of the PDF.  The point estimates presented are the peaks of the PDFs.

\begin{figure*}
\includegraphics[width=\columnwidth]{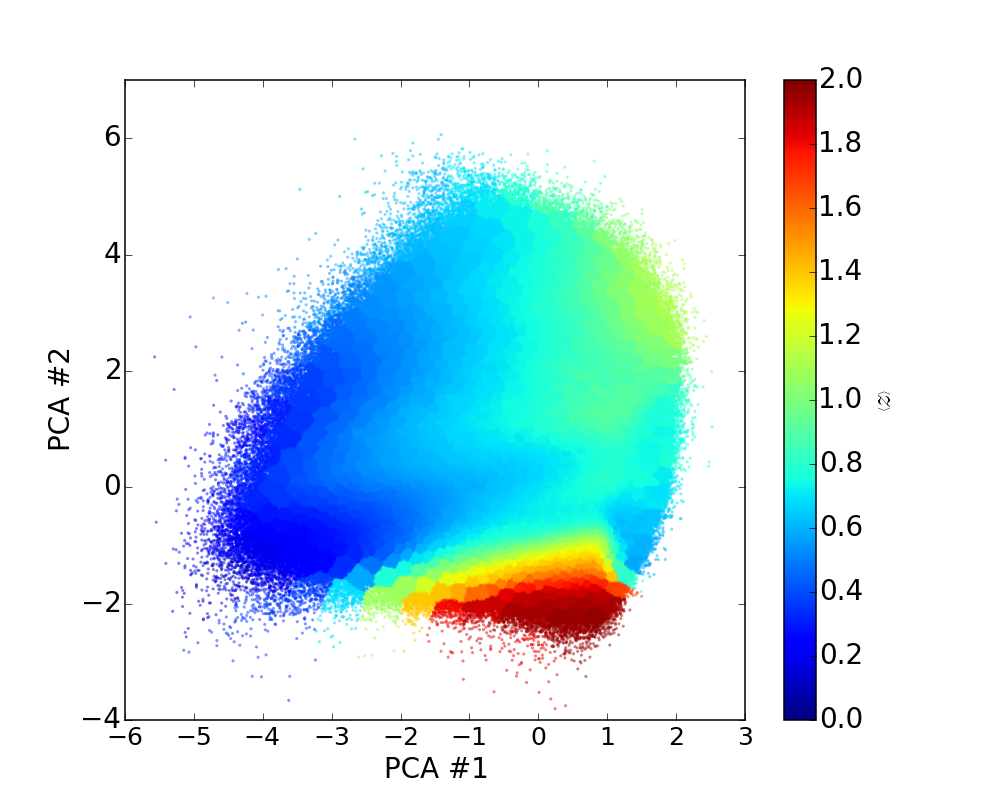}
\includegraphics[width=\columnwidth]{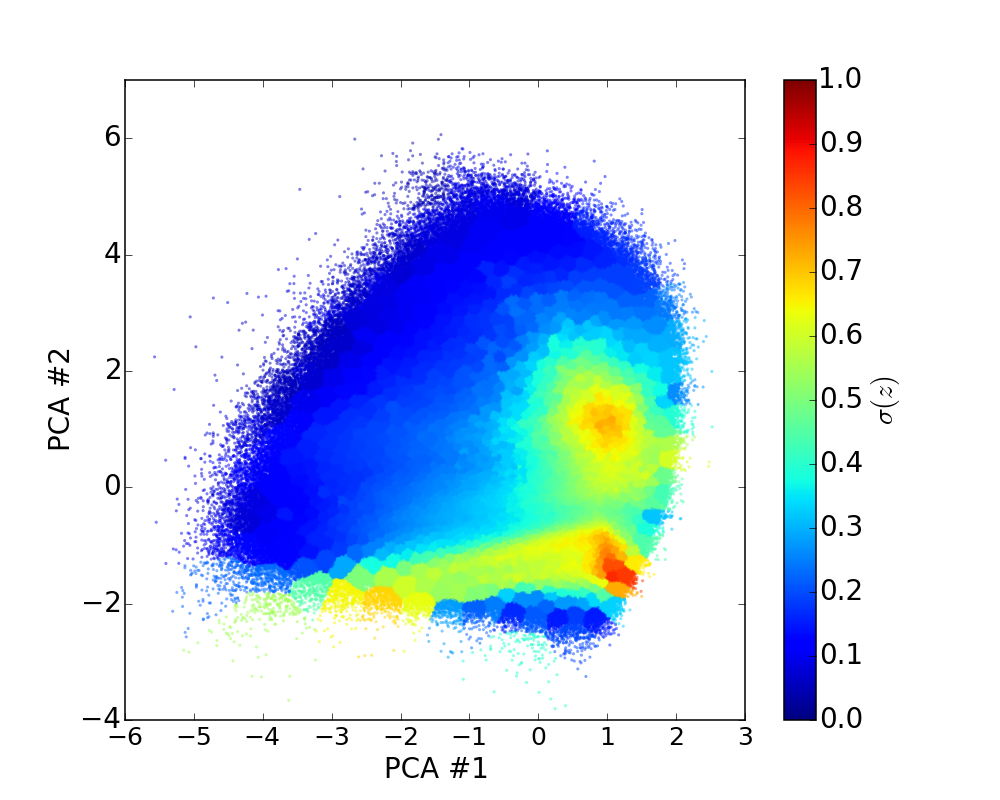}
\caption{Left: the mean redshift $\avg{z}$ in each of the 1700 Voronoi cells identified by k-means algorithm from the PCA projected data. Right: the standard deviation of the redshift $\sigma(z)$ of the galaxies in each of the 1700 Voronoi cells identified by k-means algorithm from the PCA projected data. A low value of PCA\#1 corresponds to high apparent brightness.}
\label{fig:kmeansRedshift}
\end{figure*}

Figure \ref{fig:kmeansRedshift} shows the mean and the standard deviation of the redshift of the galaxies in each of the $1700$
cells. The figure shows that the highest and the lowest redshift galaxies are separated in different regions in the plane,
the lowest average redshift cells favouring the most negative PCA\#1 values, while the highest redshift cells can be
found with the smallest PCA\#2 values. Over a large number of cells the variance between the photometric redshifts
of the galaxies in each cell is fairly low, as shown by the right panel of Figure \ref{fig:kmeansRedshift}.

\subsection{Galaxy Properties}\label{ss:galaxyProperties}

In the previous sections we have looked at the average properties of the galaxies in the Voronoi cells,
which are relevant for weak gravitational lensing. However, because of the clear trends in the average
ellipticities as shown by the left panel of Figure \ref{fig:kmeansEllipticity} it is interesting to
look also at the astrophysical properties of these galaxies. There are many parameters that describe
galaxy morphology and astrophysical properties, here we chose one that was derived from the CFHTLenS
ellipticity measurements (using \emph{lens}fit; the bulge-to-disc ratio) and one that was derived from photometric
measurements (the stellar mass) for comparison.
However this approach could be generalised to any property listed in a catalogue.

\subsubsection{Bulge-to-Disc Ratios}

\begin{figure*}
\includegraphics[width=\columnwidth]{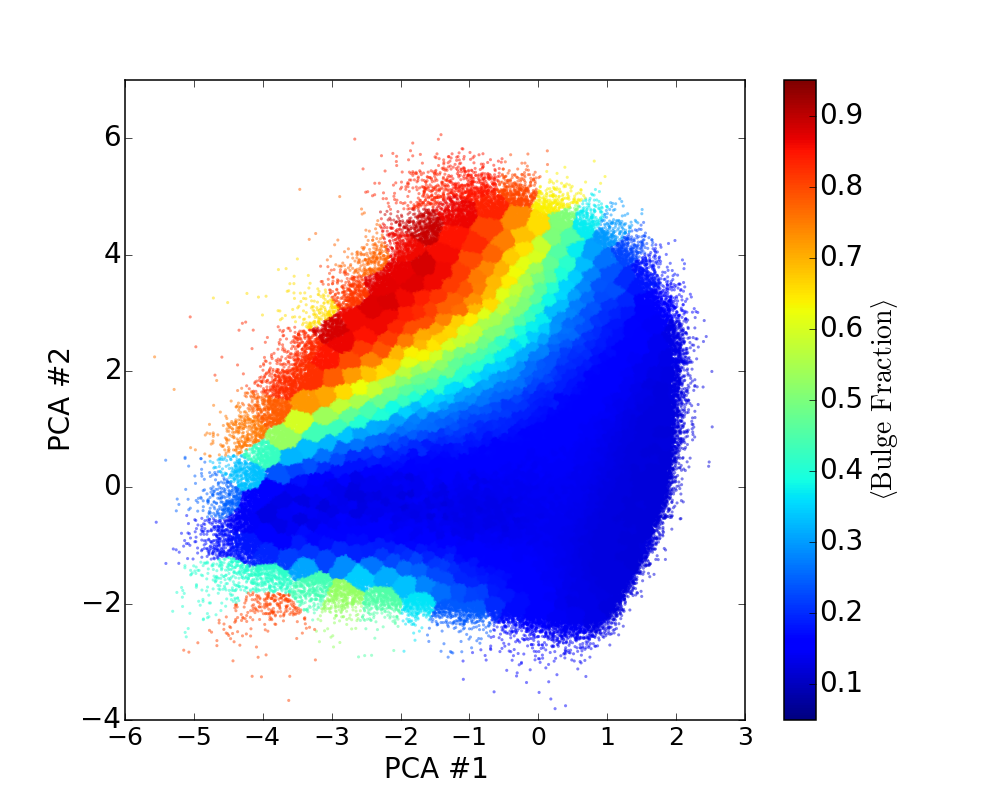}
\includegraphics[width=\columnwidth]{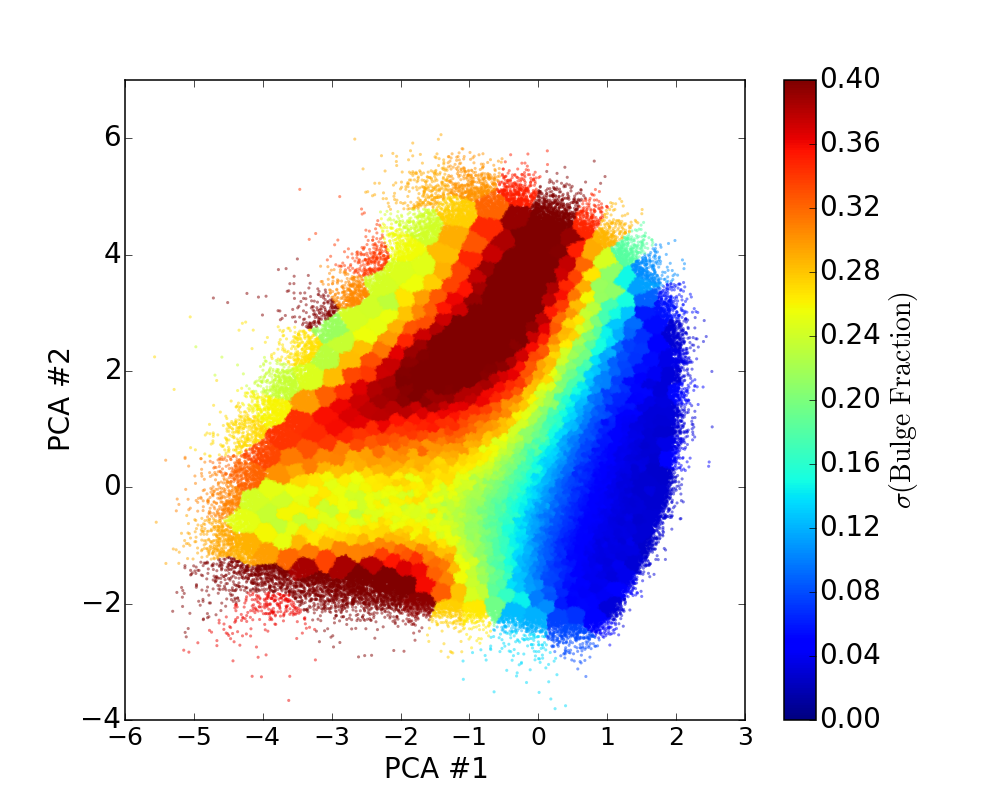}
\caption{Left: the mean bulge fraction in each of the 1700 Voronoi cells identified by k-means algorithm from the PCA projected data. Right: the standard deviation of the bulge fraction of the galaxies in each of the 1700 Voronoi cells identified by k-means algorithm from the PCA projected data. A low value of PCA\#1 corresponds to high apparent brightness.}
\label{fig:kmeansBF}
\end{figure*}

A fundamental galaxy property is the bulge-to-disc ratio, which can be used as a proxy for the morphological type of a
galaxy \citep[e.g.][]{1986ApJ...302..564S}. Figure \ref{fig:kmeansBF} shows the mean and the standard deviation of the
bulge fractions of the galaxies measured in the CFHTLenS (column \textit{BULGE\_FRACTION} in the CFHTLenS catalogue)
in each of the $1700$ k-means clusters. A clear trend can be observed; the smaller the PCA\#1 and higher
the PCA\#2 value, the higher the average bulge fraction. Thus, galaxies in these Voronoi cells are more
likely to be bulge dominated elliptical galaxies. If this is true, there should also be a similar trend in the average stellar masses.

\subsubsection{Stellar Masses}

\begin{figure*}
\includegraphics[width=\columnwidth]{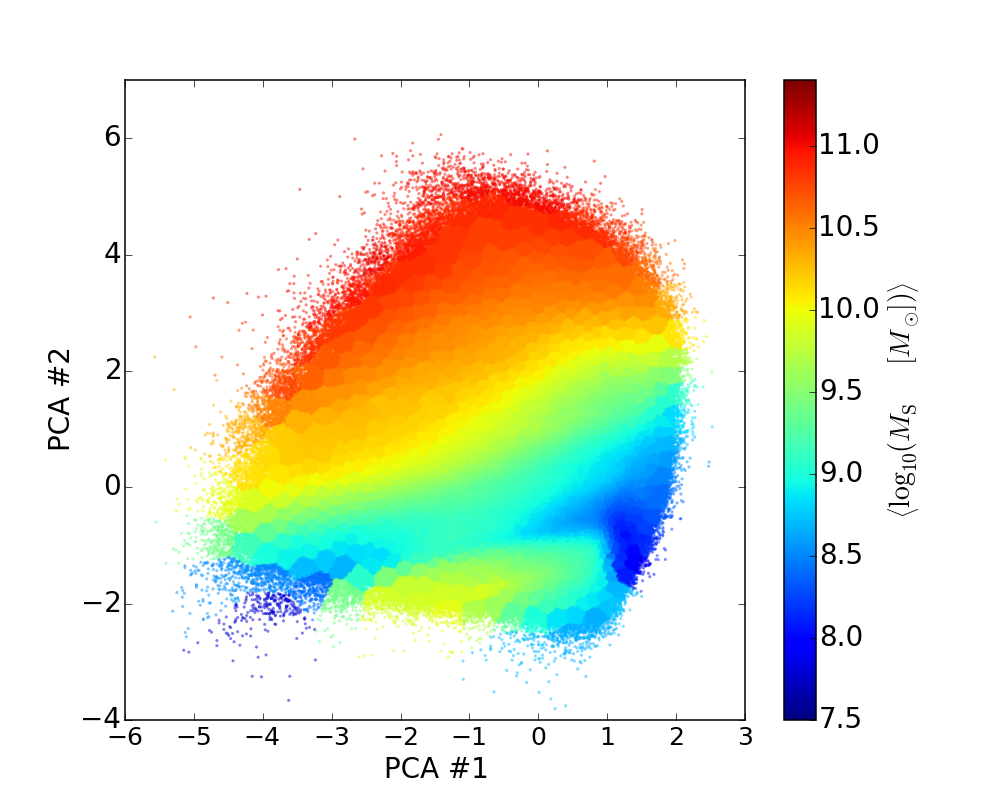}
\includegraphics[width=\columnwidth]{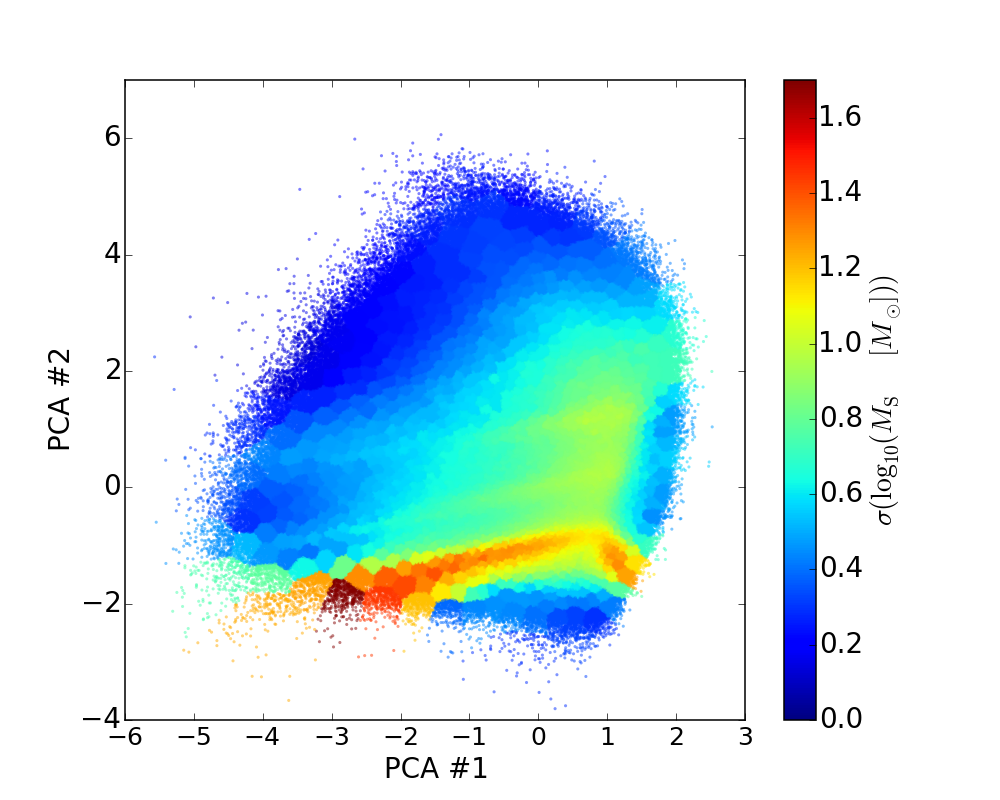}
\caption{Left: the mean stellar mass $\avg{\log_{10}(M_{S} \, [M_{\odot}])}$ in each of the 1700 Voronoi cells identified by k-means algorithm from the PCA projected data. Right: the standard deviation of the stellar mass $\sigma(\log_{10}(M_{S} \, [M_{\odot}]))$ of the galaxies in each of the 1700 Voronoi cells identified by k-means algorithm from the PCA projected data. A low value of PCA\#1 corresponds to high apparent brightness.}
\label{fig:kmeansSM}
\end{figure*}

For the stellar masses we use the \textit{LP\_log10\_SM\_MED} column of the CFHTLenS catalogue, which contains the
median estimate for the stellar mass. The missing values, indicated by $-99$, were omitted from the calculations.
Figure \ref{fig:kmeansSM} shows the mean and the standard deviation of the stellar masses of the galaxies in each
of the $1700$ k-means Voronoi cells. A clear trend along the PCA\#2 can be observed: higher values of the PCA component
correspond to average higher stellar masses, which is as expected because stellar mass is related to galaxy colour.
This agrees with the interpretation of the bulge-to-disc ratio results
for the location of elliptical galaxies in the PCA plane.

\subsection{Regions of Interest}\label{ss:regionsInterest}

The three regions, "A", "B", and "C", identified earlier can now be linked to different galaxy types. For example, the Voronoi cells in region "A", that is the highest $(\gs 3.5)$ PCA\#2 and intermediate $(\ls 0.5)$ PCA\#1 values, contain very massive elliptical galaxies, which reside on average at an intermediate redshifts $(\avg{z} \sim 0.7)$. Region "B", with the smallest PCA values (PCA\#1 $\ls -3.5$ and PCA\#2 $\ls 0.2$), contains cells with mostly relatively low mass (see Fig. \ref{fig:kmeansSM}) face-on disk or $S0$-galaxies (Fig. \ref{fig:kmeansBF}) who reside relative nearby (low redshift, see Fig. \ref{fig:kmeansRedshift}). The highest redshift $(\avg{z} \sim 2)$ disk galaxies, which are also often inclined, can be found in region "C" with the smallest $(\ls -1.5)$ PCA\#2 and intermediate $(-2 \ls \textrm{PCA\#1} \ls 1)$ PCA\#1 values. 

\section{Application: Improved Parameter Inference}\label{s:Results}

\subsection{Context}

Once the distribution functions of the statistical properties of parameters of interest are available as a function
of meta-parameters, these can then be used to improve the inferences made on model parameters that one may wish to fit to the data. This approach is tied very closely to the approach of Bayesian Hierarchical modelling \citep[e.g.][]{2014arXiv1411.2608S}, and indeed the distribution functions uncovered using our approach can be used as initial input into a Bayesian network where prior probability distributions of meta-parameters and parameters of interest are required in order to efficiently sample the joint likelihood space investigated. When including a full model of the data in such an approach, the aggregated gains commonly lead to improved parameter precision and accuracy (as has been seen in a supernovae application in cosmology, see e.g. \cite{2009ApJ...704..629M, 2011ApJ...731..120M}).

In this paper we apply the distribution of the standard deviation of ellipticity in a more simple, but also pedagogical, manner where we extend the weak lensing power spectrum formalism to include such a distribution and assess the change in expected
cosmological parameter performance derived from such an analysis.

\subsection{Generalised Cosmic Shear Power Spectra}\label{ss:FoM}

Previous cosmic shear analyses use a single population of galaxies (or split into two samples e.g. \cite{2013MNRAS.432.2433H}), after selections are made at the shape measurement stage, and a single ellipticity variance calculated and applied to calculate a single cosmic shear power spectrum. This would correspond to a delta-function distribution in the right panel of Figure \ref{fig:PDFs}. In our case the data can be subdivided into several sub-populations, each with a different number density $n_{\alpha}(z)$ and ellipticity variance $\sigma^2_{\alpha}(|e|)$ (henceforth shortened to $\sigma^{2}_{e,\alpha}$), these can be combined in a more optimal way.

The observed tomographic cosmic shear power spectrum \citep[see][]{1999ApJ...522L..21H} can be written as a sum of a signal and noise term
\be
C_{ij,\ell}=C^S_{ij,\ell}+N_{ij,\ell} \quad ,
\ee
where $ij$ refer to redshift bin pairs (such that $i=j$ refers to intra-bin power or `auto-correlation', and $i\not=j$ refers to inter-bin power or `cross-correlation'), and $\ell$ refers to an angular wavenumber. We refer the reader to \cite{2011MNRAS.413.2923K} where this is derived from a spherical harmonic-spherical Bessel representation of the full three-dimensional shear field.

The signal part of the power spectrum $C^S$ is a function of angular diameter distances of the source redshift, and
a function of the matter power spectrum as a function of redshift (or comoving distance):
\be
C_{ij,\ell}^{S} = A\int_0^{r(z_H)}{\rm d}r' \frac{W_i(r')W_j(r')}{a(r')^2}P(\ell/r'; r[z]) \quad .
\ee
The constant $A=(3\Omega_{\rm M}H_0^2/2c^2)^2$, where $\Omega_{\rm M}$ is the dimensionless matter density, $H_0$ is the current value of the Hubble parameter, $z_H$ is the redshift of the cosmic horizon, $c$ is the speed of light in a vacuum, $a(r)$ is the dimensionless scale factor, $P(\ell/r; r(z))$ is the power spectrum of matter perturbations, and $r(z)$ is a comoving distance.
The weight function is
\be
\label{weight}
W_i(r)=\int_r^{r(z_H)}{\rm d}{r'} p(r'|r[z_i])\frac{f_K(r'-r)}{f_K(r')} \quad ,
\ee
where $f_K(r)=\sinh(r)$, $r$, $\sin(r)$ for curvatures of $K=-1$, $0$, $1$, and $p(r'\, |\, r)$ is the probability that a galaxy with comoving distance $r$ is observed at distance $r'$. This representation of the shear power spectrum assumes the Limber approximation, a spherical Bessel transform to comoving distance, and a binning in redshift; for a derivation of this from the full 3D cosmic shear power spectrum see \cite{2011MNRAS.413.2923K}. 

The noise part of the power spectrum is
\be
N_{ij, \ell}=\frac{\sigma_e^2\delta^K_{ij}}{\bar N_{ii}} \quad ,
\ee
where $\sigma^{2}_{e}$ is the ellipticity variance, $\delta^K$ is a Kroneger delta-function, and $\bar N_{ii}$ is the mean number of galaxies in redshift bin $i$. To minimise the noise contribution, one should aim to maximise the mean number of galaxies, while minimising the ellipticity variance.

For each $N_{\rm bin}$ galaxy populations, each with a different number density and ellipticity variance, derived using the process described in the previous Section, the power spectra can be combined using an inverse-variance weighted sum
\be
\overline C_{ij,\ell}=\frac{\sum_{\alpha}^{N_{\rm bin}} (1/\sigma^2_{e,\alpha}) C^{\alpha}_{ij,\ell}}{\sum_{\alpha}^{N_{\rm bin}} (1/\sigma^2_{e,\alpha})} \quad ,
\ee
where $C^{\alpha}_{ij,\ell}$ is the power spectrum computed for the population $\alpha$, that has a
number density $n_{\alpha}(z)$ and an ellipticity variance $\sigma^2_{e,\alpha}$. We assume here that each population is
subdivided into the same redshift bin set, although this can in principle also change between the populations.

\subsection{Applying the Inverse Weighting}

Here we present a prediction for how such weighting will impact cosmological parameter estimation using tomographic cosmic shear.
To compute expected cosmological parameter errors we use the Fisher matrix formalism present in \cite{1999ApJ...522L..21H} for cosmic shear tomography. This results in a matrix $F_{\alpha\beta}$ (the Greek letters denote cosmological parameters) where
the $[(F^{-1})_{\alpha\alpha}]^{1/2}$ is a vector of expected, marginalised, cosmological parameter uncertainties.
We use a Cold Dark Matter (CDM) cosmology with a varying dark energy equation of state, where the free parameters that we use are
$\Omega_{\rm M}$, $\Omega_{\rm B}$, $\sigma_8$, $w_0$, $w_a$, $h$, $n_s$ (respectively, the dimensionless matter density; dimensionless baryon density; the amplitude of matter fluctuations on $8$Mpc scales -- a normalisation of the power spectrum of matter perturbations; the dark energy equation of state parametrised by $w(z)=w_0+w_a z/(1+z)$; the Hubble parameter $h=H_0/100$kms$^{-1}$Mpc$^{-1}$; and the scalar spectral index of initial matter perturbations). For each parameter we use the \textit{Planck} maximum likelihood values \citep{2014A&A...571A..15P, 2014A&A...571A..16P} about which to take derivative of the power spectra for the Fisher matrix. All parameter errors and biases we quote are marginalised over all other parameters in this set. We use the {\tt camb sources} code \citep[see \url{http://camb.info/sources/} and][]{Seljak:1996is, Lewis:2002ah, Challinor:2011bk} to compute the cosmic shear tomographic power spectra\footnote{Code to reproduce the results of this paper is available on request.}. We use a maximum radial wavenumber of $k_{\rm max}=10h$Mpc$^{-1}$ and a corresponding redshift-dependent maximum $\ell$-mode of $\ell=k_{\rm max}r[z]$.

The cosmic shear survey we assume is a {\it Euclid}-like experiment that has an area of
$15000$ square degrees, a median redshift of $z_m=0.9$, a number density of $30$ galaxies per
square arcminute with a number density distribution $n(z)$ given in \cite{2007MNRAS.374.1377T}, and
a photometric redshift distribution that is assumed to be Gaussian with a standard deviation of $\sigma(z)=0.05(1+z)$.
These characteristic are described in \cite{2011arXiv1110.3193L}.

To investigate the expected effect we first look at some simple model examples to illustrate the essential aspects of
the information change that occurs when including a more general distribution of the $p(\sigma)$. We use the
probability distribution of the variance shown in the right panel of Figure \ref{fig:PDFs}, and assume that
the number density distribution as a function of redshift is the same for each population. We compare with
the standard case, where the mean of the variance is taken from the distribution in the right panel of
Figure \ref{fig:PDFs} and a single population is assumed. We find that the inclusion of the inverse-variance weighting, in this case, improves the predicted dark energy Figure-of-Merit ($1/(\sigma(w_0)\sigma(w_a)-[\sigma^2(w_0,w_a)])$) \citep{2006astro.ph..9591A} by a modest two per cent. However, this improvement demonstrates the principle of the improvement that may be gained, and this will be more significant when several such meta-parameters of the data are combined.

\begin{figure}
\includegraphics[width=75mm]{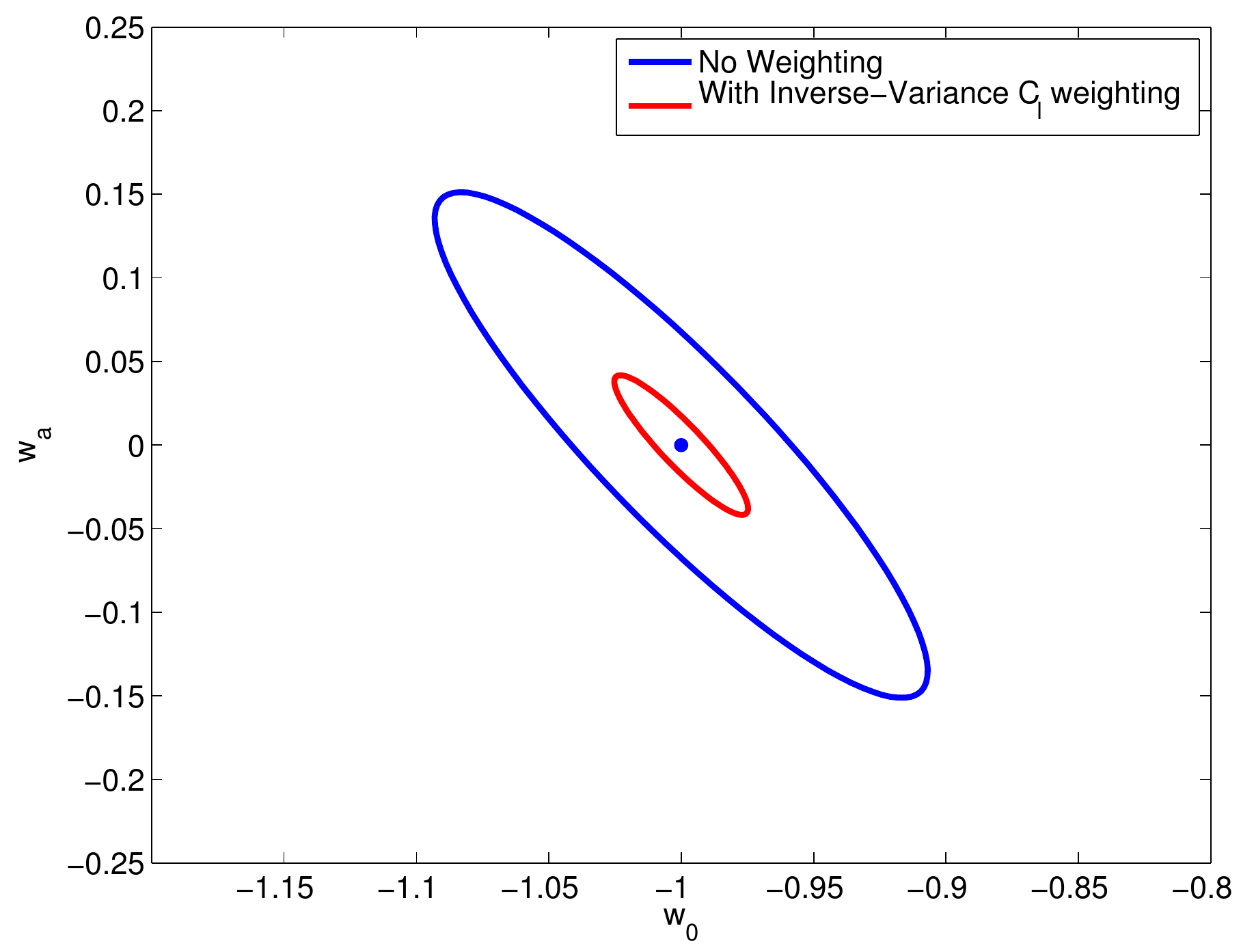}
\caption{The predicted marginalised 2-parameter $1\sigma$ uncertainty ellipsoids in the $(w_0,w_a)$ plane with and without the inverse-variance weighting, for a \Euclid-like experiment in case half of the galaxies have half the ellipticity standard deviation.}
\label{fig:w0wa}
\end{figure}

To demonstrate the principle further, we show in Figure \ref{fig:w0wa} the $(w_0,w_a)$ predicted projected $1\sigma$ uncertainty contours for the case in which no $\sigma_e$ weighting is used and the case in which half of the galaxies have a $\sigma_e$ which is half the nominal case i.e. the distribution is a top hat function from $0.15$ to $0.30$. It is clear from this Figure that if additional meta-parameters can be found, such that Voronoi cells with small ellipticity variance can be identified, significant gains can be made, and this would in fact become the dominant consideration in supplementing weak lensing surveys for dark energy science. Using supplementary data such as that discussed in Huff et al. (2013) there
remains the possibility of finding such a low-$\sigma_e$ measure or population of galaxies.


\section{Conclusion}\label{s:Conclusion}

In this paper we present a method that can be used to find how the probability distributions of parameters of interest
depend on other, meta-parameters in the data. This uses a dimensionality reduction and noise whitening, followed by a clustering
analysis. We use this to investigate how the probability distributions of galaxy ellipticity depend on photometry of galaxies.
In particular we construct a probability distribution for the intrinsic standard deviation of the galaxy ellipticities $p(\sigma(|e|))$.

We apply this to the investigation of galaxy properties and find that such an approach is particularly good at isolating galaxy type: where regions in the clustering analysis correspond to early and late type galaxies. 
Finally, we apply this to the case of improving cosmological parameter inference using weak lensing and find a few percent improvement in dark energy measurements even in this simple application.

The method we use here has several applications as a tool for cosmology, and has synergies with a full Bayesian hierarchical modelling approach.

\vspace{0.5cm}
\noindent{\em Acknowledgements:} The authors thank the anonymous reviewer whose comments helped to improve the manuscript. TDK is supported by a Royal Society University Research Fellowship. We thank the CFHTLenS team for making their catalogues public. We also thank Jason McEwen and Ignacio Ferreras for useful discussions.


\footnotesize{
\bibliographystyle{mn2e}
\bibliography{biblio}

\begin{thebibliography}{}
\makeatletter
\relax
\def\mn@urlcharsother{\let\do\@makeother \do\$\do\&\do\#\do\^\do\_\do\%\do\~}
\def\mn@doi{\begingroup\mn@urlcharsother \@ifnextchar [ {\mn@doi@}
  {\mn@doi@[]}}
\def\mn@doi@[#1]#2{\def\@tempa{#1}\ifx\@tempa\@empty \href
  {http://dx.doi.org/#2} {doi:#2}\else \href {http://dx.doi.org/#2} {#1}\fi
  \endgroup}
\def\mn@eprint#1#2{\mn@eprint@#1:#2::\@nil}
\def\mn@eprint@arXiv#1{\href {http://arxiv.org/abs/#1} {{\tt arXiv:#1}}}
\def\mn@eprint@dblp#1{\href {http://dblp.uni-trier.de/rec/bibtex/#1.xml}
  {dblp:#1}}
\def\mn@eprint@#1:#2:#3:#4\@nil{\def\@tempa {#1}\def\@tempb {#2}\def\@tempc
  {#3}\ifx \@tempc \@empty \let \@tempc \@tempb \let \@tempb \@tempa \fi \ifx
  \@tempb \@empty \def\@tempb {arXiv}\fi \@ifundefined
  {mn@eprint@\@tempb}{\@tempb:\@tempc}{\expandafter \expandafter \csname
  mn@eprint@\@tempb\endcsname \expandafter{\@tempc}}}

\bibitem[\protect\citeauthoryear{{Albrecht} et~al.,}{{Albrecht}
  et~al.}{2006}]{2006astro.ph..9591A}
{Albrecht} A.,  et~al., 2006, ArXiv Astrophysics e-prints, \href
  {http://adsabs.harvard.edu/abs/2006astro.ph..9591A} {}

\bibitem[\protect\citeauthoryear{{Amendola} et~al.,}{{Amendola}
  et~al.}{2013}]{2013LRR....16....6A}
{Amendola} L.,  et~al., 2013, \mn@doi [Living Reviews in Relativity]
  {10.12942/lrr-2013-6}, \href
  {http://adsabs.harvard.edu/abs/2013LRR....16....6A} {16, 6}

\bibitem[\protect\citeauthoryear{{Bartelmann} \& {Schneider}}{{Bartelmann} \&
  {Schneider}}{2001}]{2001PhR...340..291B}
{Bartelmann} M.,  {Schneider} P.,  2001, \mn@doi [\physrep]
  {10.1016/S0370-1573(00)00082-X}, \href
  {http://adsabs.harvard.edu/abs/2001PhR...340..291B} {340, 291}

\bibitem[\protect\citeauthoryear{{Bengio}, {Courville}  \& {Vincent}}{{Bengio}
  et~al.}{2012}]{2012arXiv1206.5538B}
{Bengio} Y.,  {Courville} A.,   {Vincent} P.,  2012, preprint, \href
  {http://adsabs.harvard.edu/abs/2012arXiv1206.5538B} {} (\mn@eprint {arXiv}
  {1206.5538})

\bibitem[\protect\citeauthoryear{{Ben{\'{\i}}tez}}{{Ben{\'{\i}}tez}}{2000}]{2000ApJ...536..571B}
{Ben{\'{\i}}tez} N.,  2000, \mn@doi [\apj] {10.1086/308947}, \href
  {http://adsabs.harvard.edu/abs/2000ApJ...536..571B} {536, 571}

\bibitem[\protect\citeauthoryear{{Benjamin} et~al.,}{{Benjamin}
  et~al.}{2007}]{2007MNRAS.381..702B}
{Benjamin} J.,  et~al., 2007, \mn@doi [\mnras]
  {10.1111/j.1365-2966.2007.12202.x}, \href
  {http://adsabs.harvard.edu/abs/2007MNRAS.381..702B} {381, 702}

\bibitem[\protect\citeauthoryear{{Bennett} et~al.,}{{Bennett}
  et~al.}{2013}]{2013ApJS..208...20B}
{Bennett} C.~L.,  et~al., 2013, \mn@doi [\apjs] {10.1088/0067-0049/208/2/20},
  \href {http://adsabs.harvard.edu/abs/2013ApJS..208...20B} {208, 20}

\bibitem[\protect\citeauthoryear{Cao, Cuevas  \& Manteiga}{Cao
  et~al.}{1994}]{Cao94}
Cao R.,  Cuevas A.,   Manteiga W.~G.,  1994, \mn@doi [Computational Statistics
  \& Data Analysis] {10.1016/0167-9473(92)00066-Z}, 17, 153

\bibitem[\protect\citeauthoryear{Challinor \& Lewis}{Challinor \&
  Lewis}{2011}]{Challinor:2011bk}
Challinor A.,  Lewis A.,  2011, \mn@doi [Phys.Rev.]
  {10.1103/PhysRevD.84.043516}, D84, 043516

\bibitem[\protect\citeauthoryear{{Clifton}, {Ferreira}, {Padilla}  \&
  {Skordis}}{{Clifton} et~al.}{2012}]{2012PhR...513....1C}
{Clifton} T.,  {Ferreira} P.~G.,  {Padilla} A.,   {Skordis} C.,  2012, \mn@doi
  [\physrep] {10.1016/j.physrep.2012.01.001}, \href
  {http://adsabs.harvard.edu/abs/2012PhR...513....1C} {513, 1}

\bibitem[\protect\citeauthoryear{{Coe}, {Ben{\'{\i}}tez}, {S{\'a}nchez}, {Jee},
  {Bouwens}  \& {Ford}}{{Coe} et~al.}{2006}]{2006AJ....132..926C}
{Coe} D.,  {Ben{\'{\i}}tez} N.,  {S{\'a}nchez} S.~F.,  {Jee} M.,  {Bouwens} R.,
    {Ford} H.,  2006, \mn@doi [\aj] {10.1086/505530}, \href
  {http://adsabs.harvard.edu/abs/2006AJ....132..926C} {132, 926}

\bibitem[\protect\citeauthoryear{{Conselice}, {Bershady}  \&
  {Jangren}}{{Conselice} et~al.}{2000}]{2000ApJ...529..886C}
{Conselice} C.~J.,  {Bershady} M.~A.,   {Jangren} A.,  2000, \mn@doi [\apj]
  {10.1086/308300}, \href {http://cdsads.u-strasbg.fr/abs/2000ApJ...529..886C}
  {529, 886}

\bibitem[\protect\citeauthoryear{Ding \& He}{Ding \& He}{2004}]{kmeans2}
Ding He X.,  2004, Proc. of Int'l Conf. Machine Learning

\bibitem[\protect\citeauthoryear{{Duncan}, {Joachimi}, {Heavens}, {Heymans}  \&
  {Hildebrandt}}{{Duncan} et~al.}{2014}]{2014MNRAS.437.2471D}
{Duncan} C.~A.~J.,  {Joachimi} B.,  {Heavens} A.~F.,  {Heymans} C.,
  {Hildebrandt} H.,  2014, \mn@doi [\mnras] {10.1093/mnras/stt2060}, \href
  {http://adsabs.harvard.edu/abs/2014MNRAS.437.2471D} {437, 2471}

\bibitem[\protect\citeauthoryear{{Erben} et~al.,}{{Erben}
  et~al.}{2013}]{2013MNRAS.433.2545E}
{Erben} T.,  et~al., 2013, \mn@doi [\mnras] {10.1093/mnras/stt928}, \href
  {http://adsabs.harvard.edu/abs/2013MNRAS.433.2545E} {433, 2545}

\bibitem[\protect\citeauthoryear{{Fu} et~al.,}{{Fu}
  et~al.}{2008}]{2008A&A...479....9F}
{Fu} L.,  et~al., 2008, \mn@doi [\aap] {10.1051/0004-6361:20078522}, \href
  {http://adsabs.harvard.edu/abs/2008A%26A...479....9F} {479, 9}

\bibitem[\protect\citeauthoryear{{Halko}, {Martinsson}  \& {Tropp}}{{Halko}
  et~al.}{2009}]{2009arXiv0909.4061H}
{Halko} N.,  {Martinsson} P.-G.,   {Tropp} J.~A.,  2009, preprint, \href
  {http://adsabs.harvard.edu/abs/2009arXiv0909.4061H} {} (\mn@eprint {arXiv}
  {0909.4061})

\bibitem[\protect\citeauthoryear{Hastie, Tibshirani  \& Friedman}{Hastie
  et~al.}{2001}]{hastie01statisticallearning}
Hastie T.,  Tibshirani R.,   Friedman J.,  2001, The Elements of Statistical
  Learning.
Springer Series in Statistics, Springer New York Inc., New York, NY, USA

\bibitem[\protect\citeauthoryear{{Heymans} et~al.,}{{Heymans}
  et~al.}{2012}]{2012MNRAS.427..146H}
{Heymans} C.,  et~al., 2012, \mn@doi [\mnras]
  {10.1111/j.1365-2966.2012.21952.x}, \href
  {http://adsabs.harvard.edu/abs/2012MNRAS.427..146H} {427, 146}

\bibitem[\protect\citeauthoryear{{Heymans} et~al.,}{{Heymans}
  et~al.}{2013}]{2013MNRAS.432.2433H}
{Heymans} C.,  et~al., 2013, \mn@doi [\mnras] {10.1093/mnras/stt601}, \href
  {http://adsabs.harvard.edu/abs/2013MNRAS.432.2433H} {432, 2433}

\bibitem[\protect\citeauthoryear{{Hildebrandt} et~al.,}{{Hildebrandt}
  et~al.}{2012}]{2012MNRAS.421.2355H}
{Hildebrandt} H.,  et~al., 2012, \mn@doi [\mnras]
  {10.1111/j.1365-2966.2012.20468.x}, \href
  {http://adsabs.harvard.edu/abs/2012MNRAS.421.2355H} {421, 2355}

\bibitem[\protect\citeauthoryear{{Hoekstra} \& {Jain}}{{Hoekstra} \&
  {Jain}}{2008}]{2008ARNPS..58...99H}
{Hoekstra} H.,  {Jain} B.,  2008, \mn@doi [Annual Review of Nuclear and
  Particle Science] {10.1146/annurev.nucl.58.110707.171151}, \href
  {http://adsabs.harvard.edu/abs/2008ARNPS..58...99H} {58, 99}

\bibitem[\protect\citeauthoryear{{Hu}}{{Hu}}{1999}]{1999ApJ...522L..21H}
{Hu} W.,  1999, \mn@doi [\apjl] {10.1086/312210}, \href
  {http://adsabs.harvard.edu/abs/1999ApJ...522L..21H} {522, L21}

\bibitem[\protect\citeauthoryear{{Huff}, {Krause}, {Eifler}, {George}  \&
  {Schlegel}}{{Huff} et~al.}{2013}]{2013arXiv1311.1489H}
{Huff} E.~M.,  {Krause} E.,  {Eifler} T.,  {George} M.~R.,   {Schlegel} D.,
  2013, preprint, \href {http://adsabs.harvard.edu/abs/2013arXiv1311.1489H} {}
  (\mn@eprint {arXiv} {1311.1489})

\bibitem[\protect\citeauthoryear{{Jaffe}, {Banday}, {Eriksen}, {G{\'o}rski}  \&
  {Hansen}}{{Jaffe} et~al.}{2005}]{2005ApJ...629L...1J}
{Jaffe} T.~R.,  {Banday} A.~J.,  {Eriksen} H.~K.,  {G{\'o}rski} K.~M.,
  {Hansen} F.~K.,  2005, \mn@doi [\apjl] {10.1086/444454}, \href
  {http://adsabs.harvard.edu/abs/2005ApJ...629L...1J} {629, L1}

\bibitem[\protect\citeauthoryear{Joachimi, Semboloni, Bett, Hartlap, Hilbert,
  Hoekstra, Schneider  \& Schrabback}{Joachimi et~al.}{2013}]{Joachimi01052013}
Joachimi B.,  Semboloni E.,  Bett P.~E.,  Hartlap J.,  Hilbert S.,  Hoekstra
  H.,  Schneider P.,   Schrabback T.,  2013, \mn@doi [Monthly Notices of the
  Royal Astronomical Society] {10.1093/mnras/stt172}, 431, 477

\bibitem[\protect\citeauthoryear{{Kaiser}}{{Kaiser}}{1992}]{1992ApJ...388..272K}
{Kaiser} N.,  1992, \mn@doi [\apj] {10.1086/171151}, \href
  {http://adsabs.harvard.edu/abs/1992ApJ...388..272K} {388, 272}

\bibitem[\protect\citeauthoryear{{Kilbinger} et~al.,}{{Kilbinger}
  et~al.}{2013}]{2013MNRAS.430.2200K}
{Kilbinger} M.,  et~al., 2013, \mn@doi [\mnras] {10.1093/mnras/stt041}, \href
  {http://adsabs.harvard.edu/abs/2013MNRAS.430.2200K} {430, 2200}

\bibitem[\protect\citeauthoryear{{Kitching}, {Heavens}, {Taylor}, {Brown},
  {Meisenheimer}, {Wolf}, {Gray}  \& {Bacon}}{{Kitching}
  et~al.}{2007}]{2007MNRAS.376..771K}
{Kitching} T.~D.,  {Heavens} A.~F.,  {Taylor} A.~N.,  {Brown} M.~L.,
  {Meisenheimer} K.,  {Wolf} C.,  {Gray} M.~E.,   {Bacon} D.~J.,  2007, \mn@doi
  [\mnras] {10.1111/j.1365-2966.2007.11473.x}, \href
  {http://adsabs.harvard.edu/abs/2007MNRAS.376..771K} {376, 771}

\bibitem[\protect\citeauthoryear{{Kitching}, {Heavens}  \& {Miller}}{{Kitching}
  et~al.}{2011}]{2011MNRAS.413.2923K}
{Kitching} T.~D.,  {Heavens} A.~F.,   {Miller} L.,  2011, \mn@doi [\mnras]
  {10.1111/j.1365-2966.2011.18369.x}, \href
  {http://adsabs.harvard.edu/abs/2011MNRAS.413.2923K} {413, 2923}

\bibitem[\protect\citeauthoryear{{Kitching} et~al.,}{{Kitching}
  et~al.}{2014}]{2014MNRAS.442.1326K}
{Kitching} T.~D.,  et~al., 2014, \mn@doi [\mnras] {10.1093/mnras/stu934}, \href
  {http://adsabs.harvard.edu/abs/2014MNRAS.442.1326K} {442, 1326}

\bibitem[\protect\citeauthoryear{Kohonen}{Kohonen}{1982}]{KohonenMap}
Kohonen T.,  1982, \mn@doi [Biological Cybernetics] {10.1007/BF00337288}, 43,
  59

\bibitem[\protect\citeauthoryear{{Komatsu} et~al.,}{{Komatsu}
  et~al.}{2011}]{2011ApJS..192...18K}
{Komatsu} E.,  et~al., 2011, \mn@doi [\apjs] {10.1088/0067-0049/192/2/18},
  \href {http://adsabs.harvard.edu/abs/2011ApJS..192...18K} {192, 18}

\bibitem[\protect\citeauthoryear{{Laureijs} et~al.,}{{Laureijs}
  et~al.}{2011}]{2011arXiv1110.3193L}
{Laureijs} R.,  et~al., 2011, preprint, \href
  {http://adsabs.harvard.edu/abs/2011arXiv1110.3193L} {} (\mn@eprint {arXiv}
  {1110.3193})

\bibitem[\protect\citeauthoryear{{Laureijs} et~al.,}{{Laureijs}
  et~al.}{2012}]{2012SPIE.8442E..0TL}
{Laureijs} R.,  et~al., 2012, in Society of Photo-Optical Instrumentation
  Engineers (SPIE) Conference Series. , \mn@doi{10.1117/12.926496}

\bibitem[\protect\citeauthoryear{Lewis \& Bridle}{Lewis \&
  Bridle}{2002}]{Lewis:2002ah}
Lewis A.,  Bridle S.,  2002, Phys. Rev., D66, 103511

\bibitem[\protect\citeauthoryear{{MacQueen}}{{MacQueen}}{1967}]{zbMATH03340881}
{MacQueen} J.,  1967, {Some methods for classification and analysis of
  multivariate observations.}, {Proc. 5th Berkeley Symp. Math. Stat. Probab.,
  Univ. Calif. 1965/66, 1, 281-297 (1967).}

\bibitem[\protect\citeauthoryear{{Mandel}, {Wood-Vasey}, {Friedman}  \&
  {Kirshner}}{{Mandel} et~al.}{2009}]{2009ApJ...704..629M}
{Mandel} K.~S.,  {Wood-Vasey} W.~M.,  {Friedman} A.~S.,   {Kirshner} R.~P.,
  2009, \mn@doi [\apj] {10.1088/0004-637X/704/1/629}, \href
  {http://adsabs.harvard.edu/abs/2009ApJ...704..629M} {704, 629}

\bibitem[\protect\citeauthoryear{{Mandel}, {Narayan}  \& {Kirshner}}{{Mandel}
  et~al.}{2011}]{2011ApJ...731..120M}
{Mandel} K.~S.,  {Narayan} G.,   {Kirshner} R.~P.,  2011, \mn@doi [\apj]
  {10.1088/0004-637X/731/2/120}, \href
  {http://adsabs.harvard.edu/abs/2011ApJ...731..120M} {731, 120}

\bibitem[\protect\citeauthoryear{{Massey}, {Kitching}  \& {Richard}}{{Massey}
  et~al.}{2010}]{2010RPPh...73h6901M}
{Massey} R.,  {Kitching} T.,   {Richard} J.,  2010, \mn@doi [Reports on
  Progress in Physics] {10.1088/0034-4885/73/8/086901}, \href
  {http://adsabs.harvard.edu/abs/2010RPPh...73h6901M} {73, 086901}

\bibitem[\protect\citeauthoryear{{McEwen}, {Josset}, {Feeney}, {Peiris}  \&
  {Lasenby}}{{McEwen} et~al.}{2013}]{2013MNRAS.436.3680M}
{McEwen} J.~D.,  {Josset} T.,  {Feeney} S.~M.,  {Peiris} H.~V.,   {Lasenby}
  A.~N.,  2013, \mn@doi [\mnras] {10.1093/mnras/stt1855}, \href
  {http://adsabs.harvard.edu/abs/2013MNRAS.436.3680M} {436, 3680}

\bibitem[\protect\citeauthoryear{Miller, Kitching, Heymans, Heavens  \&
  Van~Waerbeke}{Miller et~al.}{2007}]{Miller21112007}
Miller L.,  Kitching T.~D.,  Heymans C.,  Heavens A.~F.,   Van~Waerbeke L.,
  2007, \mn@doi [Monthly Notices of the Royal Astronomical Society]
  {10.1111/j.1365-2966.2007.12363.x}, 382, 315

\bibitem[\protect\citeauthoryear{Miller et~al.,}{Miller
  et~al.}{2013}]{Miller11032013}
Miller L.,  et~al., 2013, \mn@doi [Monthly Notices of the Royal Astronomical
  Society] {10.1093/mnras/sts454}, 429, 2858

\bibitem[\protect\citeauthoryear{{Munshi}, {Valageas}, {van Waerbeke}  \&
  {Heavens}}{{Munshi} et~al.}{2008}]{2008PhR...462...67M}
{Munshi} D.,  {Valageas} P.,  {van Waerbeke} L.,   {Heavens} A.,  2008, \mn@doi
  [\physrep] {10.1016/j.physrep.2008.02.003}, \href
  {http://adsabs.harvard.edu/abs/2008PhR...462...67M} {462, 67}

\bibitem[\protect\citeauthoryear{Parzen}{Parzen}{1962}]{parzen1962}
Parzen E.,  1962, \mn@doi [Ann. Math. Statist.] {10.1214/aoms/1177704472}, 33,
  1065

\bibitem[\protect\citeauthoryear{{Peacock}, {Schneider}, {Efstathiou}, {Ellis},
  {Leibundgut}, {Lilly}  \& {Mellier}}{{Peacock}
  et~al.}{2006}]{2006ewg3.rept.....P}
{Peacock} J.~A.,  {Schneider} P.,  {Efstathiou} G.,  {Ellis} J.~R.,
  {Leibundgut} B.,  {Lilly} S.~J.,   {Mellier} Y.,  2006, Technical report,
  {ESA-ESO Working Group on ''Fundamental Cosmology''}.
 (\mn@eprint {} {astro-ph/0610906})

\bibitem[\protect\citeauthoryear{Pearson}{Pearson}{1901}]{doi:10.1080/14786440109462720}
Pearson K.,  1901, \mn@doi [Philosophical Magazine Series 6]
  {10.1080/14786440109462720}, 2, 559

\bibitem[\protect\citeauthoryear{Pedregosa et~al.,}{Pedregosa
  et~al.}{2011}]{scikit-learn}
Pedregosa F.,  et~al., 2011, Journal of Machine Learning Research, 12, 2825

\bibitem[\protect\citeauthoryear{{Perlmutter} et~al.,}{{Perlmutter}
  et~al.}{1999}]{1999ApJ...517..565P}
{Perlmutter} S.,  et~al., 1999, \mn@doi [\apj] {10.1086/307221}, \href
  {http://adsabs.harvard.edu/abs/1999ApJ...517..565P} {517, 565}

\bibitem[\protect\citeauthoryear{{Planck Collaboration}}{{Planck
  Collaboration}}{2015}]{2015arXiv150201589P}
{Planck Collaboration} 2015, preprint, \href
  {http://adsabs.harvard.edu/abs/2015arXiv150201589P} {} (\mn@eprint {arXiv}
  {1502.01589})

\bibitem[\protect\citeauthoryear{{Planck Collaboration} et~al.,}{{Planck
  Collaboration} et~al.}{2014a}]{2014A&A...571A..15P}
{Planck Collaboration} et~al., 2014a, \mn@doi [\aap]
  {10.1051/0004-6361/201321573}, \href
  {http://adsabs.harvard.edu/abs/2014A%26A...571A..15P} {571, A15}

\bibitem[\protect\citeauthoryear{{Planck Collaboration} et~al.,}{{Planck
  Collaboration} et~al.}{2014b}]{2014A&A...571A..16P}
{Planck Collaboration} et~al., 2014b, \mn@doi [\aap]
  {10.1051/0004-6361/201321591}, \href
  {http://adsabs.harvard.edu/abs/2014A%26A...571A..16P} {571, A16}

\bibitem[\protect\citeauthoryear{{Planck Collaboration} et~al.,}{{Planck
  Collaboration} et~al.}{2014c}]{2014A&A...571A..26P}
{Planck Collaboration} et~al., 2014c, \mn@doi [\aap]
  {10.1051/0004-6361/201321546}, \href
  {http://adsabs.harvard.edu/abs/2014A%26A...571A..26P} {571, A26}

\bibitem[\protect\citeauthoryear{{Planck Collaboration} et~al.,}{{Planck
  Collaboration} et~al.}{2015}]{2015arXiv150201593P}
{Planck Collaboration} et~al., 2015, preprint, \href
  {http://adsabs.harvard.edu/abs/2015arXiv150201593P} {} (\mn@eprint {arXiv}
  {1502.01593})

\bibitem[\protect\citeauthoryear{{Riess} et~al.,}{{Riess}
  et~al.}{1998}]{1998AJ....116.1009R}
{Riess} A.~G.,  et~al., 1998, \mn@doi [\aj] {10.1086/300499}, \href
  {http://adsabs.harvard.edu/abs/1998AJ....116.1009R} {116, 1009}

\bibitem[\protect\citeauthoryear{Rosenblatt}{Rosenblatt}{1956}]{rosenblatt1956}
Rosenblatt M.,  1956, \mn@doi [Ann. Math. Statist.] {10.1214/aoms/1177728190},
  27, 832

\bibitem[\protect\citeauthoryear{Rudemo}{Rudemo}{1982}]{Rudemo1982}
Rudemo M.,  1982, Scandinavian Journal of Statistics, 9, pp. 65

\bibitem[\protect\citeauthoryear{{Schneider}, {Hogg}, {Marshall}, {Dawson},
  {Meyers}, {Bard}  \& {Lang}}{{Schneider} et~al.}{2014}]{2014arXiv1411.2608S}
{Schneider} M.~D.,  {Hogg} D.~W.,  {Marshall} P.~J.,  {Dawson} W.~A.,  {Meyers}
  J.,  {Bard} D.~J.,   {Lang} D.,  2014, preprint, \href
  {http://adsabs.harvard.edu/abs/2014arXiv1411.2608S} {} (\mn@eprint {arXiv}
  {1411.2608})

\bibitem[\protect\citeauthoryear{{Schrabback} et~al.,}{{Schrabback}
  et~al.}{2010}]{2010A&A...516A..63S}
{Schrabback} T.,  et~al., 2010, \mn@doi [\aap] {10.1051/0004-6361/200913577},
  \href {http://adsabs.harvard.edu/abs/2010A%26A...516A..63S} {516, A63}

\bibitem[\protect\citeauthoryear{Seljak \& Zaldarriaga}{Seljak \&
  Zaldarriaga}{1996}]{Seljak:1996is}
Seljak U.,  Zaldarriaga M.,  1996, Astrophys. J., 469, 437

\bibitem[\protect\citeauthoryear{{Semboloni} et~al.,}{{Semboloni}
  et~al.}{2006}]{2006A&A...452...51S}
{Semboloni} E.,  et~al., 2006, \mn@doi [\aap] {10.1051/0004-6361:20054479},
  \href {http://adsabs.harvard.edu/abs/2006A%26A...452...51S} {452, 51}

\bibitem[\protect\citeauthoryear{{Simien} \& {de Vaucouleurs}}{{Simien} \& {de
  Vaucouleurs}}{1986}]{1986ApJ...302..564S}
{Simien} F.,  {de Vaucouleurs} G.,  1986, \mn@doi [\apj] {10.1086/164015},
  \href {http://adsabs.harvard.edu/abs/1986ApJ...302..564S} {302, 564}

\bibitem[\protect\citeauthoryear{{Steinhaus}}{{Steinhaus}}{1957}]{zbMATH03129892}
{Steinhaus} H.,  1957, {Bull. Acad. Pol. Sci., Cl. III}, 4, 801

\bibitem[\protect\citeauthoryear{{Taylor}, {Kitching}, {Bacon}  \&
  {Heavens}}{{Taylor} et~al.}{2007}]{2007MNRAS.374.1377T}
{Taylor} A.~N.,  {Kitching} T.~D.,  {Bacon} D.~J.,   {Heavens} A.~F.,  2007,
  \mn@doi [\mnras] {10.1111/j.1365-2966.2006.11257.x}, \href
  {http://adsabs.harvard.edu/abs/2007MNRAS.374.1377T} {374, 1377}

\bibitem[\protect\citeauthoryear{{Viola}, {Kitching}  \& {Joachimi}}{{Viola}
  et~al.}{2014}]{2014MNRAS.439.1909V}
{Viola} M.,  {Kitching} T.~D.,   {Joachimi} B.,  2014, \mn@doi [\mnras]
  {10.1093/mnras/stu071}, \href
  {http://adsabs.harvard.edu/abs/2014MNRAS.439.1909V} {439, 1909}

\bibitem[\protect\citeauthoryear{{Weinberg}, {Mortonson}, {Eisenstein},
  {Hirata}, {Riess}  \& {Rozo}}{{Weinberg} et~al.}{2013}]{2013PhR...530...87W}
{Weinberg} D.~H.,  {Mortonson} M.~J.,  {Eisenstein} D.~J.,  {Hirata} C.,
  {Riess} A.~G.,   {Rozo} E.,  2013, \mn@doi [\physrep]
  {10.1016/j.physrep.2013.05.001}, \href
  {http://adsabs.harvard.edu/abs/2013PhR...530...87W} {530, 87}

\bibitem[\protect\citeauthoryear{Zha, Ding, Gu, He  \& Simon}{Zha
  et~al.}{2001}]{kmeans1}
Zha H.,  Ding C.,  Gu M.,  He X.,   Simon H.,  2001, Neural Information
  Processing Systems, 14

\bibitem[\protect\citeauthoryear{{van den Bergh}, {Abraham}, {Ellis}, {Tanvir},
  {Santiago}  \& {Glazebrook}}{{van den Bergh}
  et~al.}{1996}]{1996AJ....112..359V}
{van den Bergh} S.,  {Abraham} R.~G.,  {Ellis} R.~S.,  {Tanvir} N.~R.,
  {Santiago} B.~X.,   {Glazebrook} K.~G.,  1996, \mn@doi [\aj]
  {10.1086/118020}, \href {http://cdsads.u-strasbg.fr/abs/1996AJ....112..359V}
  {112, 359}

\makeatother
\end{thebibliography}
}


\onecolumn

\end{document}